\documentclass[structabstract]{aa}
\usepackage{graphicx}
\usepackage{txfonts}
\usepackage{longtable}

\def\ltsim{\raise 2pt \hbox {$<$} \kern-1.1em \lower 4pt \hbox {$\sim$}}
\def\gtsim{\raise 2pt \hbox {$>$} \kern-1.1em \lower 4pt \hbox {$\sim$}}

\begin{document}

\title{The Bologna complete sample of nearby radio sources}
\subtitle{II -- phase referenced observations of faint nuclear sources}

\author{E. Liuzzo\inst{1,2}, G. Giovannini\inst{1,2}, 
M. Giroletti\inst{1}, G.B. Taylor\inst{3,4}}
 
\institute{(1)INAF Istituto di Radioastronomia, via Gobetti 101, 40129
Bologna, Italy \\
           (2) Dipartimento di Astronomia, Universit\`a di Bologna, via 
Ranzani 1, 40127 Bologna, Italy \\ 
           (3) Department of Physics and Astronomy, University of New 
Mexico, Albuquerque NM 87131, USA \\
           (4) also Adjunct Astronomer at the National Radio Astronomy
Observatory, USA   }

\abstract
  {To study statistical properties of different classes of sources, it is
necessary to observe a sample that is free of selection
effects.}
  {To do this, we initiated a
project to observe a complete sample of radio galaxies selected from
the B2 Catalogue of Radio Sources and the Third Cambridge Revised
Catalogue (3CR), with no selection constraint
on the nuclear properties. We named this sample ``the Bologna Complete 
Sample'' (BCS).}
  {We present new VLBI observations at 5 and 1.6 GHz for 33 sources
  drawn from a sample not biased toward orientation.  By combining these data
  with those in the literature, information on the parsec-scale
  morphology is available for a total of 76 of 94 radio sources with a
  range in radio power and kiloparsec-scale morphologies.}  
{The fraction of two-sided sources at milliarcsecond resolution is high
  (30$\%$), compared to the fraction found in VLBI surveys selected at
  centimeter wavelengths, as expected from the predictions of unified
  models. The parsec-scale jets are generally found to be straight
  and to line up with the kiloparsec-scale jets.  A few peculiar
  sources are discussed in detail.}
{}


\maketitle

\section{Introduction}

The study of the parsec scale properties of radio galaxies is crucial
for obtaining information on the nature of their central engine.  
To study the statistical properties of different classes of sources, it is
necessary to define and observe a sample that is free of selection
effects. To this aim, it is important to select samples using low
radio frequencies. Sources in low-frequency samples are dominated 
by their extended and unbeamed (isotropic) emission, rather than the beamed 
compact emission that dominates in high-frequency surveys. Low 
frequency surveys are therefore unbiased with respect to the 
orientation of the nuclear relativistic jet. 
With this purpose in mind, we initiated a
project to observe a complete sample of radio galaxies selected from
the B2 Catalogue of Radio Sources and the Third Cambridge Revised
Catalogue (3CR) (\cite{g90, Giovannini2001}), with no selection constraint
on the nuclear properties. We named this sample ``the Bologna Complete
Sample'' (BCS).

In the original sample, 95 radio sources from the B2 and 3CR catalogues were
present, but because of the rejection of one source discussed here, we
redefined the complete sample to be 94 sources. We selected
the sources to be stronger than a flux density limit of 
0.25 Jy at 408 MHz for 
the B2 sources and greater than 10 Jy at 178 MHz for the 3CR sources (Feretti 
et al. 1984) and applied the following criteria:\\
1) declination $>10^{\circ}$;\\
2) Galactic latitude $|b|>15^{\circ}$;\\
3) redshift z$<$0.1\\ 

In paper I (\cite{Giovannini2005}), we presented the sample and discussed
VLBI observations for an initial group of 53 sources having bright nuclear
emission. Observations were carried on with the standard VLBI technique.
We found that:

\begin{itemize}

\item No intrinsic difference in apparent jet speed and morphology on the
  parsec scale has been found between high and low power radio
  galaxies. A one-sided jet morphology is the predominant structure on
  the parsec scale, however $\sim$30$\%$ of the observed sources show
  evidence of a two-sided structure which has been quite rare in
  previous observations of the sources with high-power cores. This
  result is in rough agreement with a random orientation of radio
  galaxies and a high jet velocity ($\beta$ $\sim$ 0.9).

\item With very few exceptions, the parsec and the kiloparsec-scale
  radio structures are aligned confirming that the large bends present
  in some BL-Lacs are likely amplified by their small angles between
  the jet direction and the line-of-sight. In sources with aligned
  parsec and kiloparsec scale structure, the main jet is always on the
  same side as the parsec-scale emission.

\item In $\sim$30$\%$ of the sources, the correlated VLBI flux density
  is less than $70\%$ of the arcsecond core flux density at the same
  frequency suggesting the presence of sub-kiloparsec-scale structures.

\item No polarized flux has been detected, confirming the low level of
  polarized emission in radio galaxies at parsec resolution.

\end{itemize}

\noindent
Moreover a few sources have been discussed in detail and results have been 
presented in several papers (references are provided to these in 
Table \ref{table1}).  

\indent Here we present new VLBA observations at 5 GHz for 26 radio
galaxies with a core flux density $S_{5GHz} > 5$ mJy, 23 of them
observed for the first time at mas resolution. We also present new
VLBA observations at 1.6 GHz for 10 radio galaxies in order to study
their extended nuclear emission.  Since a few sources were observed at
both frequencies, the total number of sources with new data presented
here is 33.  For all sources we report on parsec-scale observations
and discuss in detail their structure and properties.

To complete VLBI observations of the BCS sample, 18
sources of the sample will be observed with VLBI in the future. Since these
sources have a very faint nuclear emission in the radio band we will need very
sensitive VLBI observations (large bandwidths, long integration times, and 
phase referencing).

We assume here a Hubble constant H$_{0}$= 70 km sec$^{-1}$Mpc$^{-1}$, 
$\Omega_{M}$=0.3 and $\Omega_{\lambda}$=0.7.

\section {The Sample}
In Table \ref{table1}, we give the complete list of radio sources and summarize
the most relevant information. We reported the IAU name (column 1),
other name (column 2) and the redshift of the sources (column 3). For
the kiloparsec morphology, we indicate if the source is a Faranoff
type I or II radio galaxy (FRI/II), or a compact source, further
distinguishing between flat spectrum compact sources (C), and Compact
Steep Spectrum sources (CSS). Among flat spectrum compact sources, two
are identified with BL-Lac type objects. Since these are well known
objects (Mkn 421 and Mkn 501), in Table \ref{table1} and in the 
text we refer to
them as BL-Lac objects. We indicate with S$_{c, 5}$ the arcsecond core
flux density at 5 GHz and with Log\,P$_{c}$ the corresponding
logarithm of the radio power. Log\,P$_{t}$ is the total radio power at
408 MHz. Notes in the last column refer to the status of VLBI
observations: (N) new VLBI data are given for the first time in this
paper, (N*) the source was previously observed with VLBI, but new data
are presented here, (G) see \cite{Giovannini2001}, (I) see paper I
(\cite{Giovannini2005}). Numbers refer to recent (post 2001) papers
where a few peculiar sources have been discussed in detail.

For all sources, high-quality images at arcsecond resolution obtained
with the Very Large Array (VLA) of the NRAO\footnote {The National
  Radio Astronomy Observatory is operated by Associated Universities,
  Inc., under cooperative agreement with the National Science
  Foundation.} are available in the literature, allowing a detailed study
of the large-scale structure. According to the kiloparsec scale
morphology, the sample contains 65 FRI radio galaxies, 16 FRII
sources, and 13 compact sources. According to the optical and
large-scale radio structure, one source (0722+30) has been classified
as a spiral galaxy and therefore should not be included in this
sample.  Because of its peculiarity it will be discussed in Giroletti
et al. (in preparation). The source is reported in Table \ref{table1}, but has
not been considered in the statistical considerations and will not be
considered in the future.

We refer to the ``arcsecond core'' for the measure of the nuclear emission
at arcsecond resolution provided by VLA observations at 5 GHz (see
e.g. \cite{g88}), while we use the term ``VLBI core'' to refer to
the unresolved compact component visible in the VLBI images at milliarcsecond
resolution.

\section{Observations and Data Reduction.}

\subsection{VLBA Observations at 5 GHz}    

We obtained Very Long Baseline Array (VLBA) observing time at 5 GHz to 
produce high-resolution images for 26 sources of the BCS. For these 
observations, we selected sources with an estimated nuclear flux density 
S$_{c}$(5.0) at arcsecond resolution $>$ 5 mJy.

Because in most cases the VLBA target source is not sufficiently strong for
fringe-fitting and self-calibration, we used the phase referencing
technique. Moreover, calibration of atmospheric effects for either imaging or
astrometric experiments can be improved by the use of multiple phase
calibrators that enable multi-parameter solutions for phase effects in the
atmosphere. We therefore adopted a multi-calibrator phase calibration. In
addition, a relatively long integration time for each source has been used to
obtain good $(u,v)$-coverage, necessary to properly map complex faint
structures.

Observations were made with the full VLBA. Each source was
observed with short scans at different hour angles to ensure good
$(u,v)$-coverage. We observed in full polarization (RCP and LCP) with
4 IFs (central frequency 4.971 GHz, 4.979 GHz, 4.987 GHz and 4.995 GHz). 
Both right- and left-circular polarizations were recorded using 2 bit 
sampling across a bandwidth of 8 MHz. The VLBA correlator produced 16 
frequency channels per IF/polarisation for a total aggregate bit
rate of 256 Mbs.  In order to improve the VLBI astrometric accuracy
and image quality of a target source, we observed more than one
reference calibrator along with the target.  We scheduled 
the observation so as to use
the AIPS task ATMCA, which combines the phase or multi-band delay
information from several calibrators.  Each pointing on a target
source was bracketed by a phase calibrator scan using a 4 minute duty
cycle (2.5 minutes on source, 1.5 minutes on phase calibrator). After
4 scans of target and phase calibrators we introduced one ATMCA
calibrator scan (1.5 minutes on each ATMCA calibrator).  The total
observing time on each target is about 110 minutes. Calibrators were
chosen from the VLBA calibrator list to be bright and close to the
source; in Table \ref{t.list}, we report the list of the selected
calibrators. Short scans on strong sources were interspersed with the
targets and the calibrators as fringe finders.  The
observations were correlated in Socorro, NM. Postcorrelation
processing used the NRAO AIPS package (\cite{co93}) and the Caltech 
Difmap
packages (\cite{pea94}). 
We produced also images of the calibrators in order to
obtain and apply more accurate phase and gain corrections to the
multi-sources file during the initial calibration steps; then we
determined the absolute position of the sources. We give the
coordinates of the core candidate of each source in Table
\ref{t.list}. Initial images were derived using phase-referencing,
then self-calibrated depending on SNR.  The signal-to-noise allowed
for phase self-calibration on a very few sources.

Final images were obtained using the AIPS and Difmap packages. We produce
images with natural and uniform weights. The important parameters for the
final images of the observed sources are summarized in Table
\ref{t.parameters}.  The VLBI core properties have been estimated by using
Modelfit in Difmap. The total VLBI flux is the sum of all fitted
components. The noise level was estimated from the final images.

\subsection{VLBA Observations at 1.6 GHz}    

Observation of 10 objects were also obtained in phase reference mode
with the VLBA on 2003 August 07 and 2003 August 30 as part of projects
BG136A, and B. Observations were performed in full polarization (RCP
and LCP) with two IFs (central frequencies 1659.49 MHz and 1667.49 MHz). 
Both right- and left-circular polarizations were recorded using 2 bit 
sampling across a bandwidth of 8 MHz. The VLBA correlator produced 16 
frequency channels per IF/polarisation for a total aggregate bit
rate of 128 Mbs.

Each pointing on a target source was bracketed by a calibrator scan in
a 5 min.\ duty cycle (3 min.\ on source, 2 min.\ on the
calibrator). Two groups of (typically) 11 duty cycles were executed
for each source at different hour angles, resulting in a total of
$\sim$66 minutes per target, with a good coverage of the $(u,
v)-$plane. Calibrators were chosen from the VLBA Calibrators List to
be bright and close to the source; a list of sources with their
relative calibrator is reported together with the 5 GHz calibrators
used in Table \ref{t.list} .

The correlation was performed in Socorro and the initial calibrations
were done within AIPS. Global fringe fitting was performed on all
calibrators and the solutions were applied to the targets using a
two-point interpolation.  We then produced images of the calibrators
in order to obtain and apply more accurate phase and gain corrections
to the sources; we also determined from preliminary maps the absolute
position of the sources, which we used thereafter.  The final
calibrated single-source datasets were exported to Difmap, for imaging
and self-calibration. We produced images with the same procedure
described for the 5 GHz data, and the relevant parameters are
summarized in Table \ref{t.parameters}.

\section{Results}

We present here new observations for 33 sources (including the spiral
galaxy 0722+30, see above). We detected all sources except 0722+30 and
1339+26 at 5 GHz, and 1448+63 (3C305) at 1.6 GHz.  Among detected
sources a few exhibit an extended structure that we will discuss in
detail below. For all the other sources we give a short note and for
all resolved sources we present an image. A summary of results are
reported in Table 3.

\subsection{Sources of special interest}
 
\subsubsection{0836+29-I (\object{4C\,29.30})} 

This source ($z=0.0647$, also referred to as \object{B2\,0836+29A}) has been often
confused in the literature with the cD galaxy B2\,0836+29 (0836+29-II, at $z =
0.079$), the brightest galaxy in Abell 690, which was studied by
\cite{Venturi1995}.

\begin{figure}
\centering
\includegraphics[width=9cm]{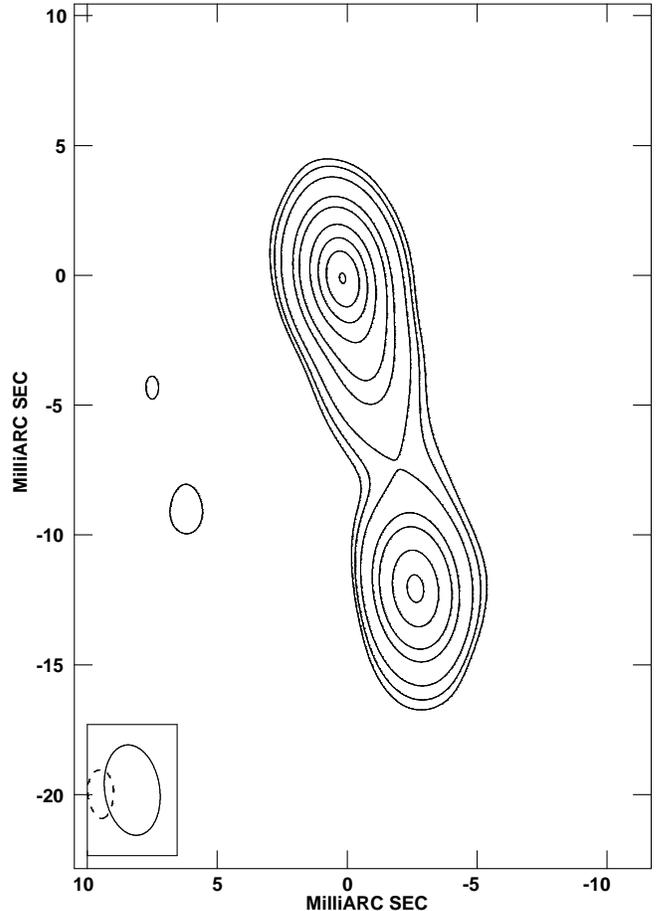}
\caption{VLBA image at 5 GHz of 0836+29-I at low resolution 
(HPBW = 35 $\times$ 21 mas in PA 8$^\circ$ ). The noise level is 0.1 mJy/beam.
Levels are $-$0.3 0.3 0.5 1 3 5 10 15 20 25 mJy/beam}
\label{f.0836a}
\end{figure}

\begin{figure}
\centering
\includegraphics[width=9cm]{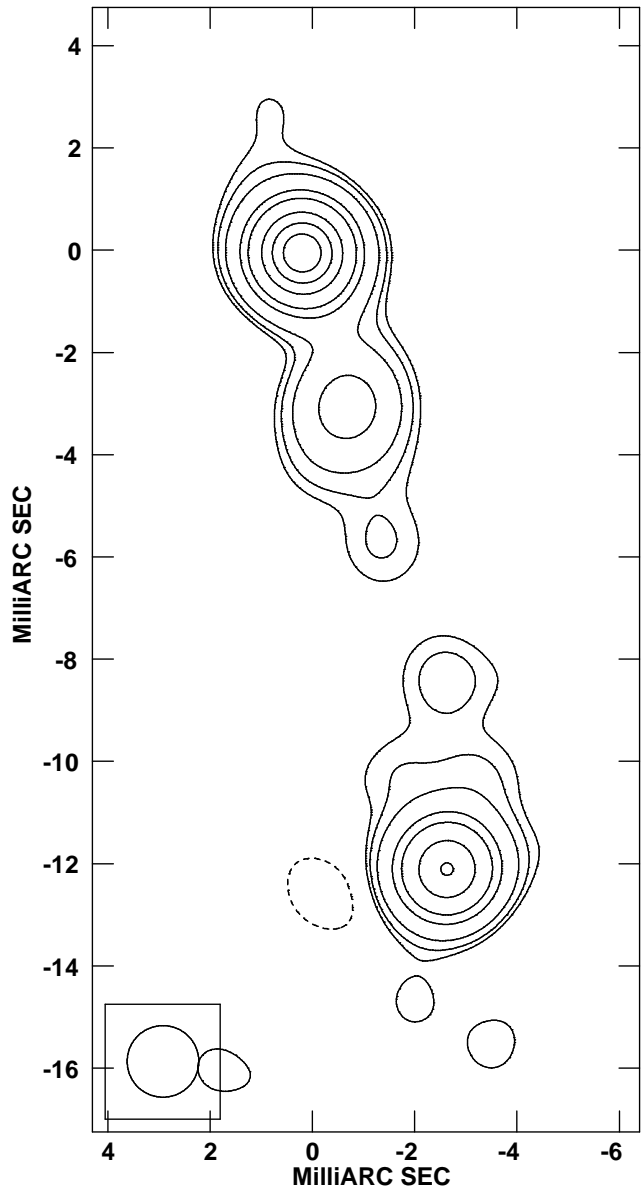}
\caption{VLBA image at 5 GHz of 0836+29-I at high resolution 
(HPBW = 1.4 mas). The noise level is 0.08 mJy/beam and contours are:
$-$0.3 0.3 0.5 1 5 10 15 20 mJy/beam. The nuclear source has been identified
with the Northern compact component (see text).}
\label{f.0836b}
\end{figure}

\object{0836+29-I} was studied in detail by \cite{breu86}, and recently by
\cite{jam07}. It is a possible merger system (\cite{breu86}).  On the
large scale the source shows clear evidence of intermittent radio 
activity. There is a large scale structure about 9' in size, with an
estimated age of about 200 Myr (\cite{jam07}); a more compact structure
shows a central core and two bright spots and an extended emission
about 1' in size with an age lower than 100 Myr. The inferred spectral
age for the inner double is 33 Myr (\cite{jam07}). At higher resolution
(\cite{breu86}) the core source (component C2) is visible along with a
bright one-sided jet about 3'' in size, which terminates with a bright
spot (C1).

In our low resolution VLBA image (Fig.~\ref{f.0836a}), 
the parsec scale
structure looks like a scaled version of the arcsecond scale structure shown by
\cite{breu86}. At higher resolution (Fig.~\ref{f.0836b}) 
the Northern 
component is resolved into a core-jet structure, and the Southern component
is resolved into a bright knot in the most external region.
Although we lack spectral information, we can identify the
nuclear source as the Northern parsec scale component, since it is the only
unresolved structure by a Gaussian fit and because of homogeneity with the
arcsecond scale.  From the jet to counter-jet ratio ($>$ 50) we estimate that
the orientation angle has to be $\theta < 49^\circ$ and the jet velocity
$\beta> 0.65$. Assuming a high jet velocity with a Lorentz factor in the range
3 to 10 (\cite{Giovannini2001}) and an orientation angle $\theta \sim 40^\circ$,
we can derive the intrinsic jet length and age of single components. In this
scenario the first knot after the core (Fig.~\ref{f.0836b}), is $\sim$ 15 
yrs old and the Southern knot is $\sim$ 70 yrs old.

The structure of this source is in agreement with a quasi-periodic
activity visible in the small as well as on the large scale.  Assuming the same
orientation angle and velocity of the jet on the arcsecond scale $\sim$ 0.6 c
(in agreement with the jet sidedness) we can estimate an age $\sim$ 10$^4$ yrs
for the one-sided jet structure present in \cite{breu86}. We note
that \cite{jam07} discussed the evidence of a strong outburst in
between 1990 and 2005, which could be related to the presence of the new VLBI
component with an estimated age of 15 yrs. In Table 1 and 4 we report the
more recent arcsecond core flux density from \cite{jam07}.
It is important to note that 
during the long time range of the activity of this source with many periods of
restarting activity the jet position angle was constant in time.

\subsubsection{\object{1350+31} (\object{3C\,293})}

This peculiar source has been studied in detail in the radio and optical
bands. Recent results are presented in \cite{Beswick2004} and
\cite{Floyd2006}, where the source structure is discussed from the sub-arcsec
to the arcminute scale. In \cite{Giovannini2005}, we presented a VLBI image at 5
GHz where a possible symmetric two-sided jet structure is present at mas
resolution. 

\begin{figure}
\centering
\includegraphics[width=9cm]{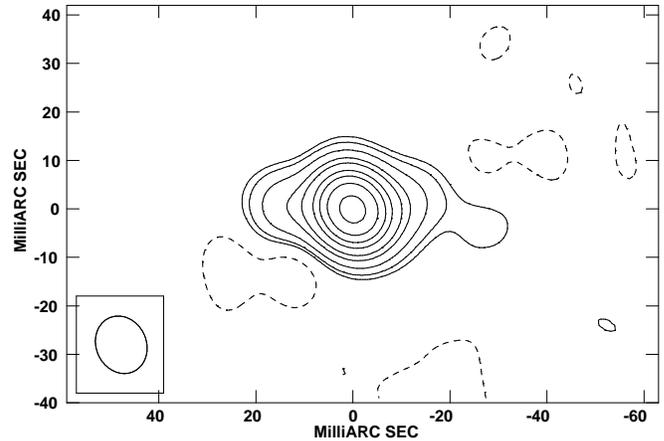}
\caption{1.6 GHz VLBA image of 1350+31 (3C\,293). The HPBW is 12 $\times$ 10 
mas in PA 24$^\circ$. The noise is 0.1 mJy/beam, contours are $-$0.3 0.3 0.5 
1 2 3 5 7 10 15 mJy/beam
\label{f1350.1}}
\end{figure}

\begin{figure}
\centering
\includegraphics[width=9cm]{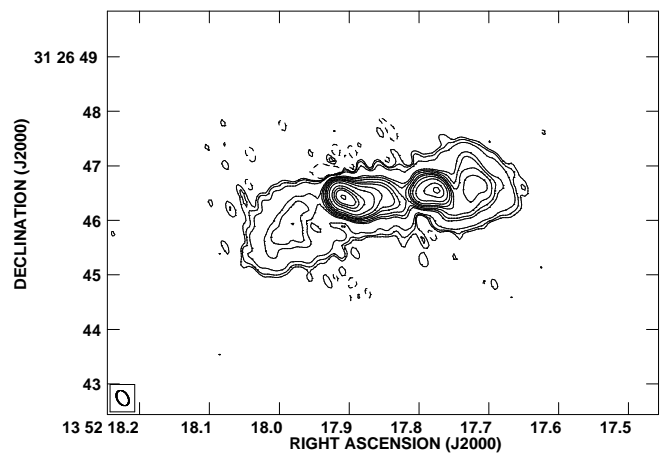}
\caption{5 GHz image of 1350+31 (3C\,293) obtained with VLA+PT. The HPBW is 
0.3'' $\times$ 0.2'' in PA 30$^\circ$. 
The noise level is 0.06 mJy/beam and levels are:
$-$0.3 0.3 0.5 1 5 7 10 15 20 30 50 100 150 200 300 mJy/beam
\label{f1350.2a}}
\end{figure}

\begin{figure}
\centering
\includegraphics[width=9cm]{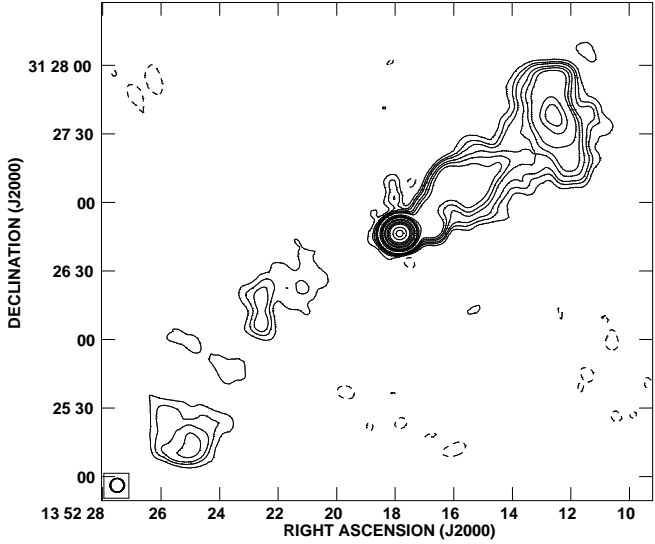}
\caption{VLA image at 1.4 GHz of 3C293. The HPBW is 6''; the noise level is 
0.26 mJy/beam, and levels are: -1 1 2 3 5 7 10 30 50 70 100 200 300 500 700 
1000 2000 3000 mJy/beam. 
\label{f1350.2b}}
\end{figure}

Because of the complex nature of the inner structure, we re-observed the
source with the VLBA at 1.6 GHz in phase reference mode. We show the new 1.6
GHz image of the core in Fig.~\ref{f1350.1}. 
In agreement with the 5 GHz VLBA
image, we have a two-sided structure with symmetric jets. The jet emission is
detected on both sides up to $\sim 20$ mas from the core, with the eastern jet
being slightly brighter. Hereafter, we refer to it as the main jet(J), and to 
the western one as the counterjet (CJ). 
This is also in agreement with the indication
discussed by \cite{Beswick2004}, although their suggested jet orientation and
velocity are in contrast with the high symmetry of the VLBA jet. From
visibility model fitting, we derive total flux densities for the core and the
jets of $S_c=17$ mJy, $S_j=2.5$ mJy, and $S_{cj}=2.0$ mJy. The core spectral
index is $\alpha_{1.6}^5=0.05$.

On the shortest baselines, we clearly detect a significant amount of extended
flux ($S\sim 200$ mJy) and we were able to image the MERLIN knots E1, E2, and 
E3 (see Fig. 3 in \cite{Beswick2004}), 
while components on the west side are not revealed (data are reported in
Table \ref{t.parameters}). We then
reduced archival VLA data in order to better study the connection between the
sub-arcsecond and the arcsecond structure. In Fig.~\ref{f1350.2a}, we
show a 5 GHz sub-arcsecond VLA+PT image. 
This reveals the inner two-sided jet
structure with a fainter diffuse emission. On even larger scales
(Fig.~\ref{f1350.2b}), VLA 1.4 GHz data reveal more extended,
fainter emission along a different PA.

>From a comparison between different images we infer that the
sub-arcsecond structure in the E-W direction is related to
restarted activity of the central AGN.  The large change in PA
with respect to the extended lobes, however, is not due to a PA change from the
restarted nuclear activity but appears constant in time and is
most likely produced by the jet interaction with a rotating disk as
discussed in \cite{breu84}. In this scenario we expect that the jet
at $\sim$ 2.5 kpc from the core is no longer relativistic.

\subsubsection{\object{1502+26} (\object{3C\,310})} 

This source is identified with a magnitude 15 elliptical galaxy at $z
= 0.0538$ having clearly extended X-ray emission, first detected with
the Einstein satellite by \cite{Burns1981}. In the radio band at low
resolution, it appears as a relaxed double \cite{breu84} with a steep
radio spectrum ($\alpha^{26}_{750}$ = 0.9; $\alpha^{0.75}_{10.7}$ =
1.4). The low brightness lobes exhibit a complex structure with filaments and
bubbles.  There is an unresolved radio core with a faint extension
from the core to the North, extended 5\arcsec, possibly identified as
a weak jet (\cite{breu84}).

\begin{figure}
\centering
\includegraphics[width=9cm]{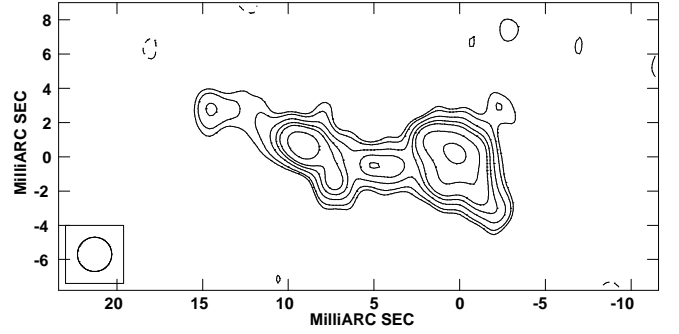}
\caption{VLBA image of 1502+26 (3C\,310) at 5 GHz. The HPBW is 2 mas. The noise
level is 0.06 mJy/beam and levels are: $-$0.2 0.15 0.2 0.3 0.4 0.5 0.7 1
mJy/beam. \label{f1502.1} }
\end{figure}

\begin{figure}
\centering
\includegraphics[width=9cm]{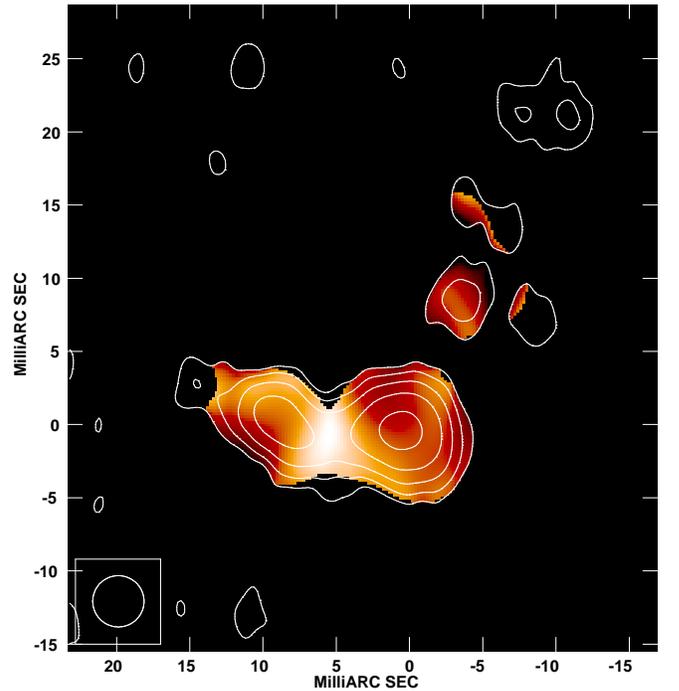}
\caption{VLBA image of 1502+26 (3C 310) at 5 GHz with a HPBW = 3.5 mas
superimposed to the spectral index image (colour) between 1.6 and 5 GHz.
The range in the colour image is from $-$0.3 (white) to 2.0 (dark red).
The noise level in the contour image is 0.04 mJy/beam and levels are: 
0.1 0.2 0.4 0.8 1.6 mJy/beam. 
\label{f1502.2} }
\end{figure}

\begin{figure}
\centering
\includegraphics[width=9cm]{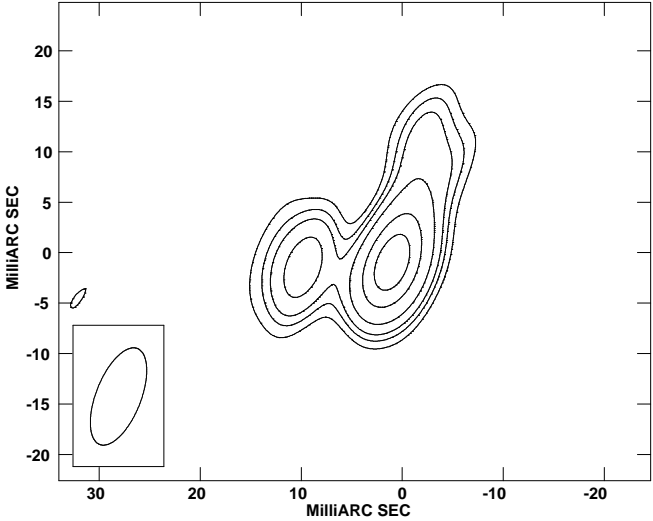}
\caption{VLBA image at 1.7 GHz of 1502+26 (3C 310), with a HPBW = 
10.2 $\times$ 4.5 mas in PA -22 $^\circ$. The noise is 0.2 mJy/beam, and
levels are: -1 1 1.5 2 3 5 7 mJy/beam.}
\label{f1502.3}
\end{figure}

The parsec scale images are complex. In \cite{Giovannini2005}, we
showed a structure interpreted as one-sided emission, in contrast with
an image by \cite{Gizani2002}. Because of this discrepancy and the
short exposure and low flux density, we reobserved this source at 5
and 1.6 GHz in phase reference mode.  In Fig.~\ref{f1502.1} we show a
5 GHz image at full resolution, where an extended complex structure
oriented in the E-W direction, almost perpendicular to the kiloparsec
scale structure is present.  The symmetry of this structure suggests
the presence of a nuclear central emission with two almost symmetrical
lobes connected by a bridge (jets?). At lower resolution
(Fig.~\ref{f1502.2}) the double E-W structure is confirmed, moreover
some emission on the top (North) of the brightest lobe is visible.  At
1.6 GHz (Fig.~\ref{f1502.3}) the source presents two main components
along the East-West direction plus a low brightness emission extending
to the North, in agreement with the low resolution 5 GHz image, and
with \cite{Gizani2002}.  In all images the source brightness is very
low, at the mJy level at 5 GHz and with a peak flux density of 8.5 mJy
at 1.6 GHz.

In order to discuss the nature of the components, we produced a map of
the spectral index (defined as $S_\nu \propto \nu^{-\alpha}$)
distribution, using images at 5 and 1.6 GHz with the same angular
resolution. The spectral index is inverted in the central region of
the E-W structure ($-$0.3) and steep in the two symmetric lobes (on
the average 0.9 at E and 1.1 at W).  Despite the low signal to noise
ratio we are able also to estimate the spectral index in the North
region which is still steep ($\sim$ 1.3), albeit with a large
uncertainty (see Fig.~\ref{f1502.2}).  Therefore we confirm the
identification of the core with the faint emission at the center of
the parsec scale structure. The morphology is quite peculiar: we have
a symmetric two-sided emission in the E-W direction, i.e. almost
perpendicular to the kiloparsec-scale emission. After about five
milliarcseconds we have a symmetric change in the direction of the
radio emission of about 80$^\circ$ which then becomes aligned with the
kiloparsec scale axis.  A possible explanation for this {\bf Z-shaped}
structure so near to the core will be discussed in Sect. 5.3.

\subsection{Notes on other observed sources}

The remaining sources presented in this paper have simpler
structures. They can be unresolved, have short jet-like features, or
even remain undetected.  Below we summarize the results on these
sources, together with some information on the large-scale
structure. We only skip 1346+26 (4C\,26.42), a very peculiar radio
source with a {\bf Z} shaped morphology from sub-parsec to kiloparsec
scales), since we will present a detailed discussion with more data in
a separate paper (\cite{liu09a}).

{\it 0034+25 (UGC367) --} This radio galaxy, identified with an E galaxy at
$z=0.032$, shows a wide angle tail morphology with an inner symmetric jet
structure in P.A.\,$=90^{\circ}$.  It belongs to the Zwicky cluster
0034.4+2532.  The host galaxy has an inner stellar disk perpendicular to the
radio jet that has been interpreted as the signature of merging in the past
(\cite{go93}).  In our VLBA observation the source appears unresolved with
total flux density $\sim$3.6 mJy.  As the ratio $S_\mathrm{VLBA}/S_{c}$ is only
36$\%$, most of the arcsecond flux is still missing on the milliarcsecond 
scale.

{\it 0055+26 (NGC326) --} This galaxy is the Brightest Cluster Galaxy (BCG) of
the Zwicky cluster 0056.9+2636. It appears as a dumbbell galaxy with two
equally bright nuclei with projected separation of 7''. The large scale
radio emission shows twin jets along direction P.A. 135$^{\circ}$ 
(\cite{Murgia2001} and references therein).  There is a peak in X-ray emission 
consistent with the location
of the radio galaxy core.  
In our VLBA image (not shown here), we found nuclear emission with a possible
slight extension in the same PA as the kiloparsec-scale jets, with a total flux
density of 7.5 mJy.

{\it 0149+35 (NGC\,708) --} This galaxy is the BCG of the cooling cluster A262.
At optical wavelengths, it is a low brightness galaxy whose nuclear regions are
crossed by an irregular dust lane and patches of dust (\cite{Capetti2000}).  
Feedback between the radio activity and the surrounding thermal gas is
discussed by \cite{Blanton2004}.  In our VLBA image, this source appears
unresolved with a total flux density $S_5 = 3.2$ mJy. Some of the arcsecond
core flux is lost on the parsec scale ($S_\mathrm{VLBA}/S_{c}\sim64\%$).  

{\it 0300+16 (3C\,76.1) --} On kiloparsec scales, 3C\,76.1 is classified as an FR I radio
galaxy; it was imaged at 6 cm by \cite{Vallee1982}, and at 20 cm by
\cite{Leahy1991}.  The source has twin jets with a faint ($\sim$ 8 mJy)
radio core surrounded by an extended low brightness radio halo.  In our VLBA
image, we found an unresolved core with the total flux density $S_5=6.1$
mJy. Some arcsecond core flux is lost on the parsec scale
(S$_\mathrm{VLBA}$/S$_{c}$$\sim$60$\%$).

{\it 0356+10 -- (3C\,98) --} The host galaxy is a typical narrow-line
radio galaxy (NLRG), with extended emission lines studied by
\cite{Baum1988,Baum1990}. On kiloparsec scales, 3C\,98 shows a
double-lobe FRII radio structure which spans 216 arcsec at 8.35 GHz,
with a radio jet that crosses the northern lobe and terminates in a
bright hotspot. There is little evidence of a southern jet, but a twin
hotspot in the southern lobe is present (\cite{Hardcastle1998}).
The core of this symmetric FRII radio galaxy is pointlike in our VLBI
image, with the total flux density $\sim$2.2 mJy. Most of the
arcsecond core flux is missing on the parsec scale
($S_\mathrm{VLBA}/S_{c}\sim24\%$), suggesting the presence of faint
jets below our noise level and invisible because they are oriented in the
plane of the sky.

{\it 0708+32B}. This source is classified as a low power compact source. In VLA images
at 20 cm (\cite{Fanti1986}) it shows a nuclear emission with two symmetric
lobes $<$ 10'' ($\sim$ 10 kpc) in size oriented in the N-S direction. 

\begin{figure}
\centering
\includegraphics[width=9cm]{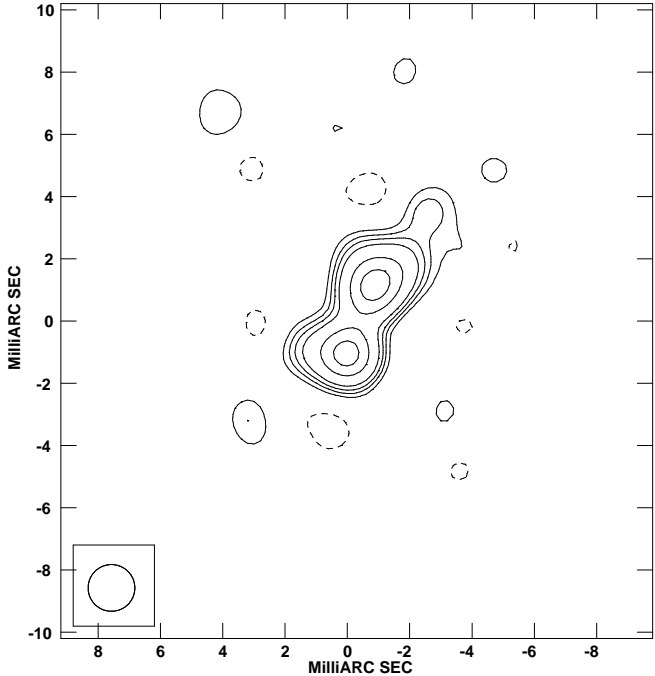}
\caption{VLBA image at 5 GHz of 0708+32B. The HPBW is 1.5 mas, the noise level
is 0.08 mJy/beam and levels are: $-$0.3 0.3 0.5 0.7 1 2 3 mJy/beam.}
\label{f0708.1}
\end{figure}

\begin{figure}
\centering
\includegraphics[width=9cm]{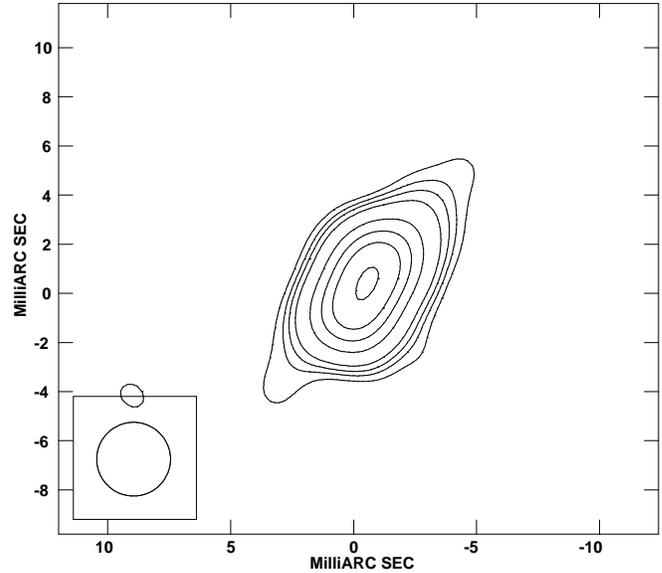}
\caption{VLBA image at 5 GHz of 0708+32B. The HPBW is 3 mas, the noise level
is 0.09 mJy/beam and levels are: $-$0.3 0.3 0.5 0.7 1 2 3 4 5 mJy/beam}
\label{f0708.2}
\end{figure}

Higher resolution VLA data will be presented in \cite{gir09} to
address the kiloparsec-scale structure of this source. In our VLBA images
(Fig.~\ref{f0708.1} and Fig.~\ref{f0708.2}), we detected a double
structure oriented along PA $\sim$ 150$^\circ$ and extended $\sim$ 4 mas.
Since no spectral index information is available we are not able to
identify the nuclear source, but the high degree of symmetry suggests 
an identification as a CSO with recurrent radio activity.  The
correlated VLBI flux density is 9.5 mJy. about 70$\%$ of the arcsecond
core flux density.

{\it 0722+30} This radio source at $z=0.0188$ is the only one identified with a
spiral galaxy (see \cite{Capetti2000}). It is a highly inclined galaxy with
a strong absorption associated with the disk and a bulge-like component. The
radio emission is quite peculiar showing two symmetric jet-like features at an
angle of about 45$^\circ$ from the disk (see \cite{Fanti1986}).  New high
resolution VLA images will be discussed in \cite{gir09}.
The source is undetected in our VLBA image, with a $S_\mathrm{peak}<0.3$
mJy/beam at 5 GHz.  Because of the optical identification this source will not
be considered in the following discussion.

{\it 0802+24 (3C\,192) --} On large scale, this radio source shows an 'X'
symmetric double-lobe structure which extends $\sim$200 arcsec at 8.35 GHz, 
showing
bright hotspots at the end of the lobes (\cite{Baum1988,Hardcastle1998}).

\begin{figure}
\centering
\includegraphics[width=9cm]{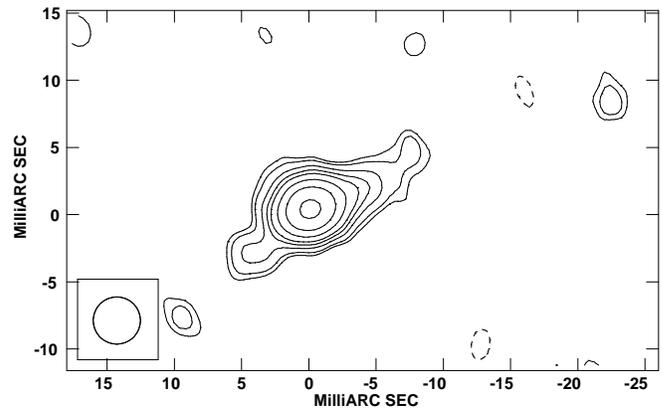}
\caption{The VLBA image at 5 GHz of 0802+24 (3C192). Contour levs are $-$0.2 0.15 
0.2 0.3 0.4 0.5 0.7 1 1.5
mJy beam$^{-1}$, and the HPBW is 3.5 mas. The noise level is 0.05 mJy/beam}
\label{f0802.1}
\end{figure}

Extended narrow line emission has been detected in this galaxy.  In
our VLBA image (Fig.~\ref{f0802.1}), the source appears two-sided, with
jets at the same orientation as the kiloparsec-scale jets 
(P.A. $\sim$-80$^{\circ}$).  The
core flux density is $\sim$2.1 mJy and the total flux density is $\sim$
3.6 mJy.  The correlated VLBI flux density is $\sim$ 50\% of the
arcsecond core flux.

{\it 0838+32 (4C\,32.26) --} This radio source is associated with a
dumbbell galaxy identified as the brightest galaxy in Abell 695,
although recent works claim that it is at the centre of a group
unrelated and much closer than the Abell cluster it was originally
identified with (\cite{Jetha2008}).  At low resolution (\cite{val79,
  Jetha2008}) it appears to be a Wide Angle Tailed (WAT) source, but at
high resolution it shows a compact FRII structure with core flux
density S$_{c, 1.4}\sim$7.5 mJy (\cite{mac98}). The radio structure
suggests a possible restarting activity and X-ray data suggest that
the currently active lobes are expanding supersonically
(\cite{Jetha2008}).  In our VLBA image, the source is unresolved with
total flux density $S_5 = 7.0$ mJy, with little or no arcsecond core
flux missing on parsec scales ($S_\mathrm{VLBA}/S_{c}\sim90\%$).

{\it 0915+32 --} The VLA radio map by \cite{Fanti1987} shows a
two-sided jet in P.A.\ $=28^{\circ}$ with mirror symmetry. The radio
structure could be due to gravitational interactions with a nearby
spiral galaxy (\cite{Parma1985}).  In our VLBA image, the source
appears unresolved with a total flux density of $S_5\sim13.6$ mJy. No
arcsecond core flux is lost on the mas scale
($S_\mathrm{VLBA}/S_{c}\sim100\%$).

{\it 1113+29 (4C\,29.41) --} This radio source is associated with the
BCG of Abell 1213, which is part of a double system of galaxies close
to the center of the cluster (\cite{Trussoni1997}). At radio
frequencies, it shows a double structure with linear size $\sim 61$
kpc and a total radio power at 1.4 GHz Log\,$P_{1.4}=24.7$
(\cite{Fanti1986}).  An extended weak X-ray emission has been found to
be associated with the cluster (\cite{Ledlow2003}).

\begin{figure}
\centering
\includegraphics[width=9cm]{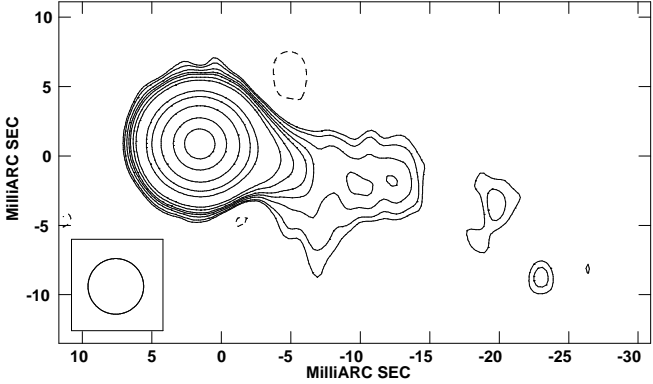}
\caption{VLBA image of 1113+29 at 5 GHz. The HPBW is 4 mas, the noise level is
0.06 mJy/beam and levels are: $-$0.2 0.15 0.2 0.3 0.4 0.5 0.7 1 3 5 10 20 30
mJy/beam}
\label{f1113.1}
\end{figure}

\begin{figure}
\centering
\includegraphics[width=9cm]{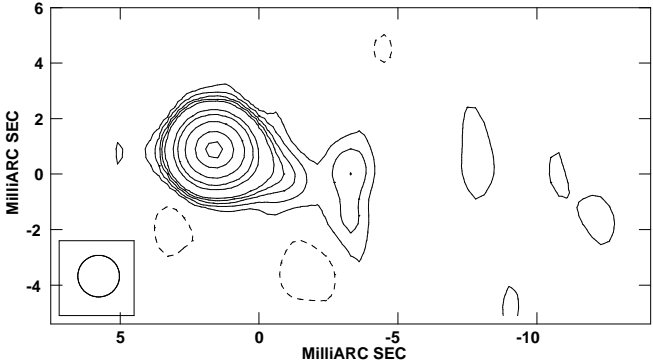}
\caption{VLBA image of 1113+29 at 5 GHz. The HPBW is 1.5 mas, the noise 
level is 0.07 mJy/beam, and levels are: $-$0.2 0.15 0.3 0.5 0.7 1 3 5 10 20 
30 mJy/beam}
\label{f1113.2}
\end{figure}
In our VLBA image, the source shows
an extended one-sided jet (Fig.~\ref{f1113.1}), resolved in the high 
resolution image and suggestive of a limb-brightened structure 
(Fig.~\ref{f1113.2}). 
The parsec-scale jet is 
on the
same side as the main kiloparsec scale jet.  The total correlated 
flux in the VLBA
image is $\sim$ 40 mJy, and therefore no arcsecond core flux is missing on the
parsec scale.

{\it 1116+28 --} The galaxy associated with this radio source is a
double system at a distance of $z = 0.0667$.  At arcsecond resolution
(\cite{Fanti1987}), the radio source shows a Narrow Angle Tail (NAT)
structure with two symmetric jets.  \cite{Giovannini2005} showed a
slightly resolved parsec-scale structure with two-sided jets oriented
in the same direction as the kpc-scale jets. Because of the large
fraction of missing flux in the VLBI image we reobserved this source
at 5 and 1.6 GHz.  We do not confirm the two-sided jets previously
reported.  We consider therefore this source to be unresolved with
total flux densities $S_{1.6}=9.5$ mJy, $S_5 =11.5$ mJy, and a
slightly inverted spectral index $\alpha_{1.6}^5=-0.17$.  Most of the
arcsecond core flux is still missing on the milliarcsecond scale at
both frequencies ($S_\mathrm{VLBA}/S_{c}\sim40\%$).

{\it 1122+39 (NGC\,3665) --} At arcsecond resolution, this appears to be
a core plus
twin-jet radio source (\cite{Parma1986}).  In our VLBA image, it shows a
pointlike structure with a total flux density $S_5 = 8.8$ mJy. No arcsecond
core flux is missing on the parcsec-scale ($S_\mathrm{VLBA}/S_{c}\sim100\%$)

{\it 1204+24 --} In the VLA image, this source appears as a FRI with faint
symmetric lobes and a bright core (\cite{Fanti1986}).  In our VLBA image, it
appears unresolved with a total flux density $S_5 = 5.3$ mJy, corresponding to
a fraction of the VLA core $S_\mathrm{VLBA}/S_{c}\sim66\%$.

{\it 1251+27B (3C\,277.3) --} The radio and optical properties of 3C\,277.3 are
described in detail by \cite{breu85}. The elliptical host galaxy of
3C\,277.3 is very smooth and regular; on kiloparsec-scales, the radio source extends $<$60 kpc and it has two wide and diffuse lobes with a few brightness
enhancements (southern jet and northern hot spot).
 

\begin{figure}
\centering
\includegraphics[width=9cm]{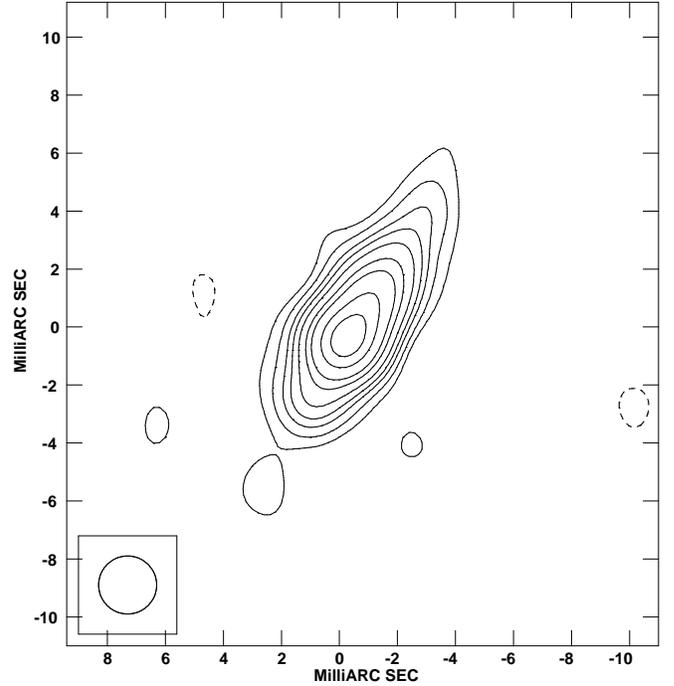}
\caption{VLBA image of 1251+27B (3C277.3). The HPBW is 2 mas, the noise is 0.06
mJy/beam, and levels are: $-$0.2 0.15 0.3 0.5 0.7 1 1.5 2 3 4 mJy/beam}
\label{f1251.1}
\end{figure}

In our VLBA image (Fig.~\ref{f1251.1}), 
this source shows a possible two-sided structure with total
flux density $\sim$10.4 mJy and a jet orientation quite similar to 
the kiloparsec-scale jet
($\Delta$PA$\sim$20$^\circ$). The core flux density is $\sim$6.2 mJy and 
only $\sim$ 10\% of
the arcsecond flux is lost on the mas scale.

{\it 1319+42 (3C\,285) --} The host galaxy of 3C\,285 has been identified with
the BCG of a group of galaxies (\cite{Sandage1972}).  3C\,285 is a
classical double-lobed radio galaxy of 190 arcsec total extension at 4.86 GHz,
with two hotspots and an eastern ridge showing curvature roughly along the line
to the optical companion (\cite{Leahy1984,Hardcastle1998}).  
\begin{figure}
\centering
\includegraphics[width=9cm]{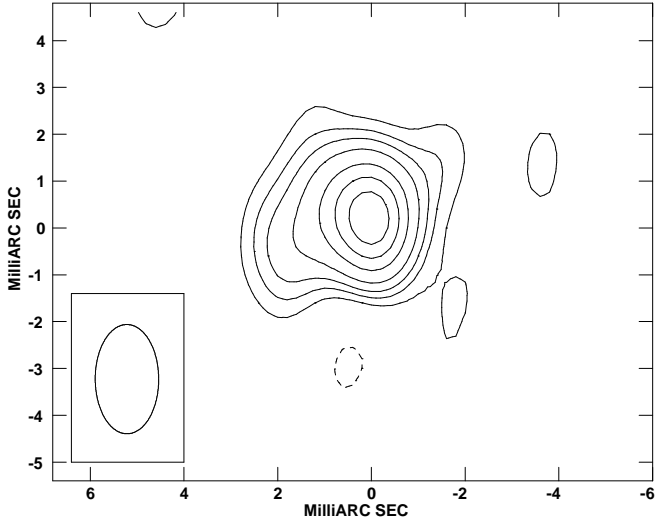}
\caption{VLBA image at 5 GHz of 1319+42 (3C 285). The HPBW is 2.3 $\times$ 1.4
mas in PA 0$^\circ$. The noise level is 0.1 mJy/beam and levels are:
$-$0.3 0.3 0.5 0.7 1 1.5 2 2.5 mJy/beam}
\label{f1319.1}
\end{figure}
In our
VLBA image, the source is slightly resolved at the same PA as the 
kiloparsec-scale
jets (Fig.~\ref{f1319.1}) with a symmetric structure. The total flux density 
is $\sim$4.7 mJy; very little arcsecond
core flux is lost on the milliarcsecond-scale (S$_\mathrm{VLBA}$/S$_{c}$$\sim$80$\%$)

{\it 1339+26 (UGC\,08669) -- } This radio galaxy has a head-tail morphology. It
is the easternmost of a double system identified with the dominant member of
the cluster Abell 1775. A small dark nuclear band (P.A.$\sim 0^{\circ}$)
characterizes this otherwise regular galaxy (\cite{Capetti2000}).  In our
VLBA image, this source is not detected above 0.3 mJy beam$^{-1}$. 

\begin{figure}
\centering
\includegraphics[width=9cm]{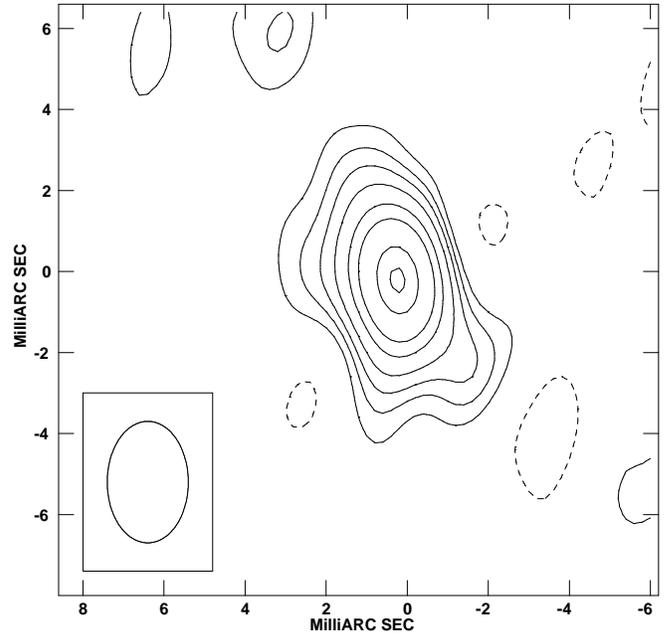}
\caption{VLBA image at 5 GHz of 1357+28. The HPBW is 3 $\times$ 2 mas in PA 
0$^\circ$. The noise is 0.15 mJy/beam, and levels are: $-$0.3 0.3 0.5 0.7 1 1.5
2 3 3.5 mJy/beam.}
\label{f1357.1}
\end{figure}

{\it 1357+28 -- } The host of this radio source is a round elliptical galaxy
with a small dust lane (P.A. $\sim 90^{\circ}$) bisecting the nuclear region
(\cite{Capetti2000}).  The radio map by \cite{Fanti1987} shows a two-sided
inner jet in P.A.  0.5$^{\circ}$ that decollimates and bends at 30 arcsec from
the core.  
Our VLBA image (Fig.~\ref{f1357.1})  
shows a slightly extended core suggesting a two-sided structure in PA $\sim$ 
20$^\circ$. The VLBA core flux
density is 5.6 mJy and some of the arcsecond core flux is missing on the milliarcsecond scale
($S_\mathrm{VLBA}/S_{c}\sim60$). 

\begin{figure}
\centering
\includegraphics[width=9cm]{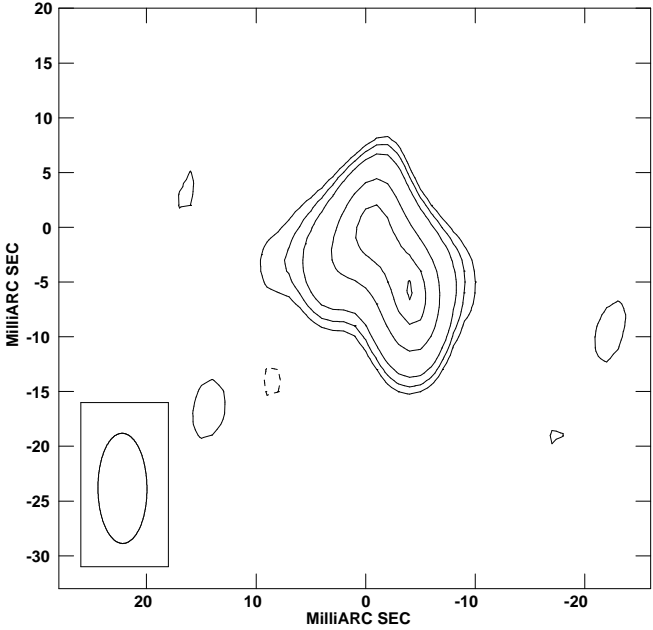}
\caption{VLBA image of 1422+26 at 1.6 GHz. The HPBW is 10.1 $\times$ 4.5 mas in
PA 0$^\circ$. The noise is 0.15 mJy/beam and levels are: $-$0.5 0.7 0.7 1 2 
3 3.5 mJy/beam.
\label{f.1422} }
\end{figure}

{\it 1422+26 -- } An extended FR I galaxy, with two symmetric jets oriented
east-west at arcsecond resolution. At 5 GHz, \cite{Giovannini2005} detect a
weak (3 mJy) milliarcsecond scale source. The 1.6 GHz observations presented 
in this work detect a larger flux
density ($S_{1.6}\sim 9.3$ mJy), distributed in a misaligned and distorted
structure (see Fig.~\ref{f.1422}).  

We tentatively identify the peak in the 1.6 GHz image with the nuclear
source detected also at 5 GHz. In this scenario the spectral index is $\sim$ 0
and the extended 1.6 GHz structure is a one-sided, distorted jet. More data
are necessary to understand the nature of this source.

{\it 1448+63 (3C\,305) -- } 
A well known double source, recently studied in detail by \cite{Massaro2009}, 
it was only marginally detected in previous VLBA 5 GHz observations (\cite{Giovannini2005}).  In our new observations at 1.6 GHz the nuclear source is not 
detected, but
there is a clear detection on the shortest
baselines. The visibilities are well described by a two component
model, separated by $\sim 1.5\arcsec$. These are probably the two hot spots
visible in the MERLIN images (see \cite{Massaro2009}).

{\it 1529+24 (3C\,321) -- } An extended FR II narrow line radio galaxy,
with optical evidence of double nuclei. The two components are clearly 
merging galaxies with evidence for a merger-triggered starburst activity 
(\cite{Roche2000}). The kiloparsec-scale nuclear emission has been discussed
by \cite{Baum1988}.

\begin{figure}
\centering
\includegraphics[width=9cm]{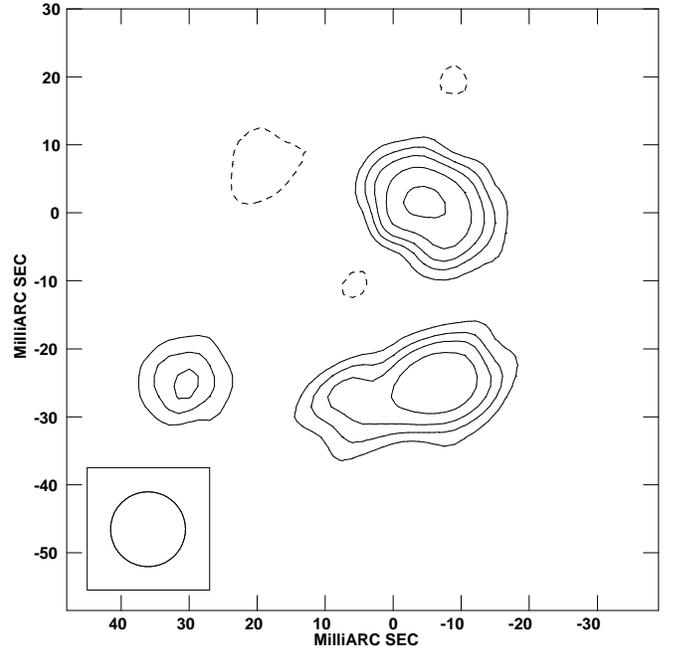}
\caption{VLBA image of 1529+24 at 1.6 GHz. The HPBW is 11.0 mas. 
The noise is 0.1 mJy/beam and levels are: $-$0.3, 0.3 0.5 0.7 1 1.5
mJy/beam.
\label{f.1529} }
\end{figure}

With the VLBA we detect at 1.6 GHz some faint, diffuse emission,
difficult
to either clean or modelfit. A tentative visibility model fit reveals $\sim 7$
mJy at 1.6 GHz, in two nearly equal components separated by $0.03\arcsec$
(see Fig.~\ref{f.1529}).  

A faint
and complex structure is expected from the large difference between the 
point-like mas
structure at 5 GHz ($\sim$ 3 mJy) and at 1.6 GHz ($\sim$ 7 mJy) and the
arcsecond core flux density at 5 GHz ($\sim$ 30 mJy).

{\it 1557+26 -- } This is an unresolved source on arcsecond scale, with $S_5=30$
mJy. Our 1.6 GHz image recovers $\sim10$ mJy in a relatively compact
component. The spectral index with respect to the 5 GHz data by
\cite{Giovannini2005} is $\alpha_{1.6}^5=0.1$ in agreement with the 
core dominated one-sided structure found at 5 GHz.
Visibility phases and
amplitudes on the LA-PT baseline suggest that there could also be a more
extended component, but it is not possible to constrain it with the present data.

{\it 1613+27 -- } A symmetric double on kiloparsec scales (\cite{Parma1986}),
with a 25 mJy core at 5 GHz. Our 1.6 VLBA image reveals an unresolved core,
with $S_\mathrm{VLBA, 1.6}=9.3$ mJy, in agreement with VLBA at 5 GHz
(\cite{Giovannini2005}); the non simultaneous spectral index is
$\alpha_{1.6}^5=0.15$. The large fraction of missing flux could imply
that the VLA core flux is overestimated, or that the source is variable.

{\it 1615+35B (NGC6109) -- }
This is a very extended Head Tail (HT) radio source with a structure 
more than 10$^{\prime}$ in size, oriented N-S. It is 
associated with an elliptical galaxy 
of m$_{pg}$=14.9 at a redshift of 0.0296. 

\begin{figure}
\centering
\includegraphics[width=9cm]{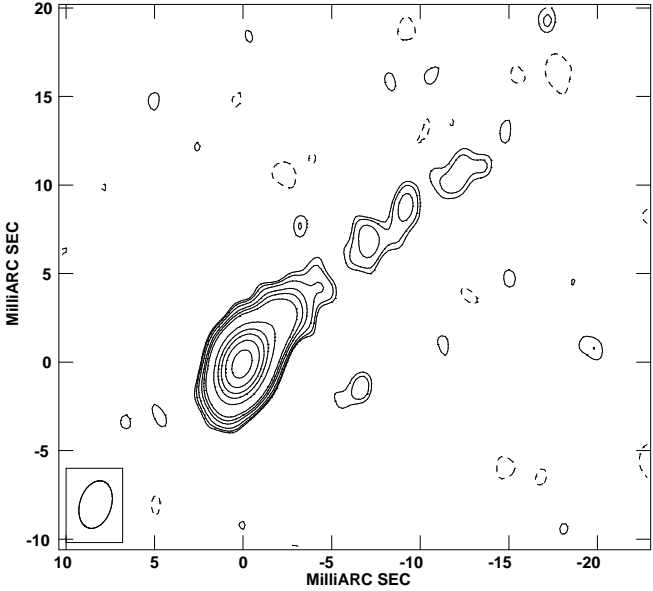}
\caption{VLBA image of 1615+35B at 5 GHz. The HBPW is 2.8 $\times$ 1.8 mas in
PA -17$^\circ$. The noise level is 0.06 mJy/beam and levels are $-$0.15 0.15
 0.2 0.3 0.5 0.7 1 3 5 7 10 15 mJy/beam.
\label{f.1615} }
\end{figure}

In our VLBA image (Fig.~\ref{f.1615}), the nuclear emission is seen 
with a jet on the same side as 
the kiloparsec-scale jet. The total flux density detected in our map is 
$\sim23.6$ mJy, the core flux is $\sim$17.1 mJy and the ratio between the 
VLBA correlated flux and the arcsecond core flux density at 5 GHz is 85$\%$. 
At 1.2 mas from the nucleus, the jet/counter-jet surface brightness ratio 
is $\geq$30. 

\begin{figure}
\centering
\includegraphics[width=9cm]{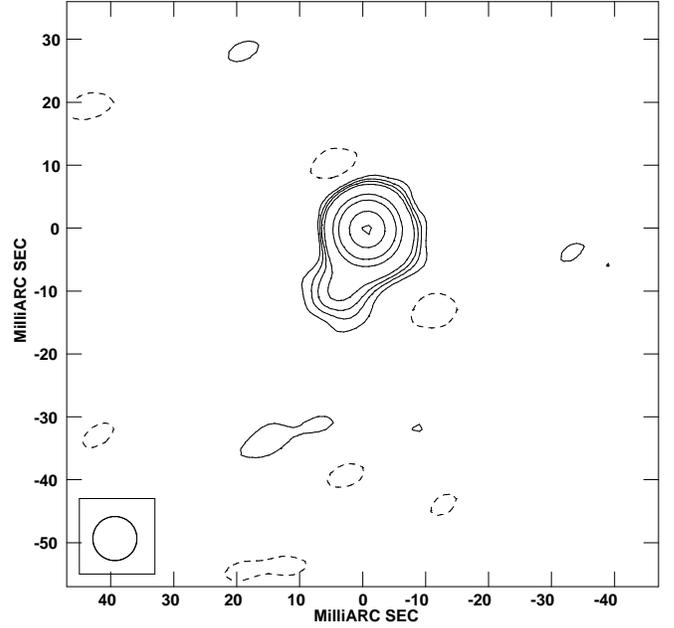}
\caption{VLBA image of 1621+38 at 1.7 GHz. The HPBW is 7 mas, the noise 0.09
mJy/beam and levels are: $-$0.3 0.3 0.5 0.7 1 3 5 10 15 mJy/beam.
\label{f.1621} }
\end{figure}

{\it 1621+38 -- } A head-tail radio galaxy (\cite{Fanti1987}). In
our VLBA 1.6 GHz data it shows 
a core-jet structure in PA $\sim 160^\circ$ (Fig.~\ref{f.1621}) in 
agreement
with the 5 GHz image of \cite{Giovannini2005}. The core spectral index is
$\alpha_{1.6}^5=0.25$ and a significant fraction of core flux density is
missing.

{\it 1637+29 -- } This peculiar source displays a head-tail morphology
associated with a poor group of galaxies (\cite{deRuiter1986}). The radio source
consists of twin bent jets, with the main jet ending in a very bright lobe and
two radio tails showing conspicuous oscillations.  Our VLBA image shows an
unresolved component with total flux density $S_5\sim =7.8$ mJy. The ratio
between the VLBA correlated flux and the arcsecond core flux density at 5 GHz
is 60$\%$.

{\it 1736+32 -- } The B2 source is resolved in two unrelated radio sources. The
main source (the Northern one) has a FR I structure, the secondary (Southern)
appears as a background FR II source projected onto the SE lobe of the FR I
galaxy (\cite{Fanti1986}).  In our VLBA image of the main source, we found an
unresolved structure with a total flux density $\sim$ 7.6 mJy. No arcsecond
core flux is missing on the milliarcsecond scale.

{\it 1752+32B}: In VLA images, this source appears as a FRI 
source (\cite{Capetti1993}).

\begin{figure}
\centering
\includegraphics[width=9cm]{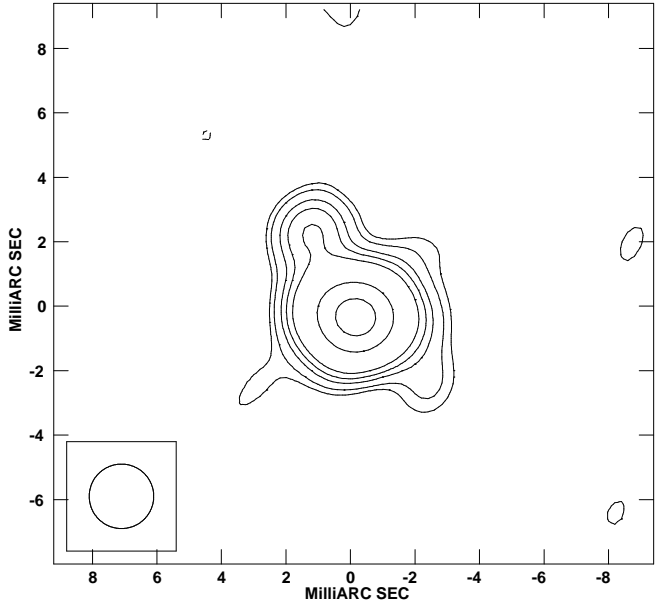}
\caption{VLBA image of 1752+32B at 5 GHz. The HPBW is 2 mas, the noise level 
0.08 mJy/beam. Contours are: $-$0.2 0.2 0.3 0.5 0.7 1 3 5 mJy/beam.
\label{f.1752} }
\end{figure}

Our VLBA image shows nuclear emission with a two-sided emission at the same 
PA as the kiloparsec-scale jet (Fig.~\ref{f.1752}). 
The total flux density detected in our map is $\sim$9.3
mJy, the core flux is $\sim$7.4 mJy and the ratio between the VLBA correlated
flux and the arcsecond core flux density at 5 GHz is 77$\%$.  

{\it 2236+35 UGC\,12127 -- } This source has a large-scale double radio jet embedded in a
low-surface-brightness region with S symmetry (\cite{Fanti1986}).  In our VLBA
image, the source is unresolved with total flux density $\sim$8.8 mJy and no
arcsecond core flux is missing on the milliarcsecond scale.

\section {Discussion}

\subsection {Source Morphology}

Among sources with VLBI data, we have the following kiloparsec-scale 
structures: (a) 51 FRI radio galaxies; (b) 13 FRII radio galaxies; (c) 11 
compact sources with a flat spectrum (including two BL-Lac objects), and 
1 compact steep-spectrum source (CSS).
The observed sample is not yet complete, but representative of a sample of 
sources 
at random angles to the line of sight. In fact, in a randomly oriented sample 
of radio galaxies, the probability that the source is at an angle 
$\theta_{1}$ and $\theta_{2}$ with respect to the line of sight is\\
\begin{center}
 P($\theta_{1}$,$\theta_{2}$) = cos$\theta_{1}$ - cos$\theta_{2}$
\end{center}
In the observed sample, the percentage of FR I plus FR II radio
sources is $84\%$ that corresponds to sources oriented at angles
greater than 33$^{\circ}$, in agreement with the result obtained in
\cite{Giovannini2005}.
We confirm with improved statistics the results discussed in \cite{Giovannini2005} and briefly review these in Sect. 1.\\
We note also that among sources that are compact at arcsecond resolution 3
(27$\%$) show a two-sided pc scale structure, 5 (45$\%$) are
one-sided, 1 (9$\%$) is a CSO and 2 (18$\%$) have not been
detected. This result confirms that low power compact sources with a
flat spectrum have sub-kpc scale structure, but their small size 
is not always due to projection effects of beamed objects as in the
case of the two BL-Lac sources and possibly the other three radio
galaxies with one-sided parsec-scale structure.

The total number of sources with a two-sided jet morphology 
on the parsec-scale is 23, 
corresponding to 30$\%$ of the sample. As discussed in
paper I (\cite{Giovannini2005}), 
this percentage is higher than that found in previous samples in the 
literature:  indeed, there are 11$\%$ (7/65) symmetric sources in the PR 
(Pearson-Readhead) sample \cite{Pearson1988} and 4.6$\%$ (19/411) in the 
combined PR and Caltech-Jodrell (CJ) samples \cite{Taylor1994, 
Polatidis1995}. The main difference between the percentage of 
symmetric sources in the present sample and in previous samples is naturally 
explained in the framework of unified scheme models by the fact that the 
present sources show relatively faint cores and they are less affected by 
orientation bias.  
For a more detailed discussion see  \cite{Giovannini2005}.\\
We find that 27$\%$ (14) of 
the FR I radio galaxies and 46$\%$ (6) of the FR II radio galaxies are 
two-sided on the parsec scale. Assuming the same intrinsic velocity, the 
higher fraction of symmetric sources among FR II galaxies could suggest that 
the jets in FR II galaxies are intrinsically brighter than those in FR I 
galaxies. No correlation is found between the jet-to-counterjet ratio and 
the total or arcsecond core radio power.

One-sided sources are found in 31/76 sources (41$\%$) confirming that this
structure is the most frequent pc scale structure because of Doppler
boosting.

The number of unresolved sources in our VLBI images
is 15 (20$\%$). Most of them (14) are FR I radio galaxies, and
1 is a FR II source (1529+24). In most cases these are sources with faint 
relativistic jets in the plane of the sky, and therefore their jets
are de-boosted.

We observed sources with a low core flux density at 5 GHz at arcsecond 
resolution ($>$ 5 mJy), but
only 3 sources (4 $\%$) have not been detected in our VLBI
observations. Two of them are
C-shaped sources as discussed before, and the other one is a FR I radio galaxy.
This result confirms that also in low power sources and in radio sources with
a low radio nuclear activity a compact nuclear radio source is present.

Finally we would like to discuss the two radio sources where
a {\bf Z}-shaped structure is present on the parsec-scale. At present
among BCS sources these are 
1502+26 (3C310), and 1346+26 (4C26.42).
A few other are present in literature as e.g. the CSO 1946+708 
(\cite{Taylor2009}).
The presence of this symmetric and large change in the jet direction so near
the core cannot be due to projection or relativistic effects. The large 
symmetry present in these sources imply that are oriented close to the plane 
of the sky, moreover a change in the jet velocity and/or orientation would
produce a change in the Doppler factor that should be revealed by the 
observations of the jet properties. As discussed in more detail in 
(\cite{liu09a}) for 1346+26 the source morphology strongly suggests that in these
sources the jets are not highly relativistic even at a few pc 
distance from the core.
A slow and possibly high density jet could more easily interact with a turbulent
surrounding medium in rotation near the central Black Hole to show this
peculiar symmetric structure. 

\subsection {Missing flux in parsec scale structures}

We compared the total flux at VLBA scale with the core arcsecond flux density 
(Table \ref{tot.parameters}). 
Over all data, among 76 sources, we find that 47 (62 $\%$) 
have a correlated flux density larger than 70$\%$ of the arcsecond core 
flux density. This means that in these sources we mapped most of the mas 
scale structure and so we can properly connect the parsec to the 
kiloparsec structures. Instead, for 19 (38$\%$) sources, a significant 
fraction of the arcsecond core flux density is missing in the VLBA images. 
This suggests variability or the presence of significant structures 
between $\sim$30 mas and 1 arcsec that the VLBA can miss due to a
lack of short baselines.  
In particular among the 6 sources where the VLBA flux
density is less than 10$\%$ of the arcsecond core flux density measured
by the VLA, we have 
evidence of a complex structure in agreement with 
an intermediate-scale structure lost in our images. Variability should be
 also considered. 
To properly study these structures, future observations with the EVLA 
at high frequency or with the 
e-MERLIN array will be necessary.

\subsection{Jet velocity and orientation}

For 51 sources we can estimate the ratio between the jet and counter-jet
brightness or provide a lower limit to the ratio (Table \ref{tot.parameters}).
Among the 25 sources with a ratio $\gtsim$ 10 we have one-sided sources
with the exception
of 0055+30 and 0220+43 where deep observations allowed us to detect a 
counterjet structure despite a large brightness ratio.
The presence of large ratios confirms the presence of relativistic jets.

We also estimated the core dominance of observed sources (Table 
\ref{tot.parameters}).
According to \cite{Giovannini1994} and references therein, a correlation is 
present between the core and total radio power and we can use the core
dominance to estimate the jet velocity and orientation.
In Table \ref{tot.parameters} we present the 
core dominance defined as the ratio between the 
observed and the estimated core radio power 
according to the relation given in \cite{Giovannini1994}.

A ratio larger than 1 implies a source with a boosted core, therefore 
oriented at an angle smaller than 60$^\circ$. For a value of the core 
dominance in the range 0.25 - 1, the core radio power is lower than the
value expected from the correlation therefore the nuclear observed power
is de-boosted and the source is oriented at an angle larger than 60$^\circ$.
These values are in general in agreement with the ratio between the jet and 
counterjet brightness and can give useful constraints on the jet velocity
and orientation.
Three sources show a core dominance larger than 10: two BL Lacs and 0222+36,
a symmetric source in VLBI images but with evidence of restarted activity
(\cite{gir05a, gir05b}).

If the core dominance is smaller than 0.25 no solution can be found from the 
correlation. Such a strong deboosting cannot be obtained even if the source 
is in the plane of the sky and the jets are strongly relativistic.
We interpret such a result as a possible nuclear variability with the core
in a low activity regime and/or the evidence that the nuclear activity
is now stopped and the total radio power is related to a previous
episode of greater nuclear activity, i.e. the AGN is dying.
We have 8 sources with a such low core dominance. Four of them are FR I 
radio galaxies (0300+16, 1122+39, 1257+28, 1357+28)
characterized by a faint nuclear emission in VLA observations
at 5 GHz not only with respect to the total low frequency flux density, but
also with respect to the source total flux density at the same frequency.
The remaining four sources are narrow line extended and powerful FR II
radio galaxies (0106+13, 0356+10, 0802+24, 1319+42), lobe dominated and
with evidence of spectral steepening at high frequency. Also in these sources
the extended radio lobes suggest a previous episode of greater 
nuclear activity.

\section {Conclusion}

In this paper, we continue the study of the BCS sample. We present here
new VLBA 
data for 33 sources, we briefly discuss the parsec and kiloparsec-scale 
structures, and we summarise the results obtained to date.

\begin{itemize}

\item The detection rate is high: only 3 sources out of 76 (4$\%$)
  have not been detected, even though we observed sources with an
  arcsecond core flux density as low as 5 mJy at 5 GHz.  This result
  confirms the presence of compact radio nuclei at the center of radio
  galaxies.

\item As expected in sources with relativistic parsec-scale jets, the
  one-sided jet morphology is the predominant structure present in our
  VLBI images, however $22\%$ of the observed sources show evidence of
  a two-sided strucure. This result is in agreement with a random
  orientation and a high jet velocity ($\beta\sim$0.9).

\item We find two sources with a {\bf Z}-shaped structure on the
  parsec-scale suggesting the presence of low velocity jets, in these
  peculiar radio sources.

\item In 8 sources the low core dominance suggests that the nuclear
  activity is now in a low activity state. The dominance of the
  extended emission implies a greater activity of the core in the
  past. However in these sources a parsec-scale core and even jets are
  present. In this scenario the nuclear activity may be in a low or
  high state but is not completely quiescent.  This result is in
  agreement with the evidence that a few sources show evidence of a
  recurring or re-starting activity. This point can be better
  addressed when observations are available for the full sample so
  that we can discuss the time-scale of the recurring activity.
 
\item In most cases, the parsec and the kiloparsec scale jet structures
  are aligned and the main jet is always on the same side with respect
  to the nuclear emission. This confirms the idea that the large bends present
  in some BL Lacs sources are amplified by the small jet orientation
  angle with respect to the line-of-sight.

\item In 62$\%$ of the sources, there is good agreement between the
  arcsecond-scale and the VLBI correlated flux density. For the other
  38$\%$ of the sources, at the milliarcsecond scale more than 30$\%$
  of the arcsecond core flux density is missing. This suggests the
  presence of variability, or of a significant sub-kiloparsec-scale
  structure, which will be better investigated with the EVLA at high
  frequency or with the e-MERLIN array.

\item To complete VLBI observations of the BCS sample, 18 sources of
  the sample will be observed with VLBI in the future. Since these
  sources have a very faint nuclear emission in the radio band we will
  require very sensitive VLBI observations (employing large bandwidths, long
  integration times, and phase referencing).
 
\end{itemize}

\begin{acknowledgements}
This research has made use of
the NASA-IPAC Extragalactic Data Base (NED) which is operated by the JPL,
California Institute of Technology, under contract with the National 
Aeronautics and Space Administration.
\end{acknowledgements}

\clearpage
\onecolumn

\begin{longtable}{cccclrcc}
\caption{The Complete Bologna Sample} \label{table1} \\
\hline\hline \\
Name & Name  & z & Morphology & S$_c$(5.0)& Log P$_c$& Log P$_t$ & Notes \\
IAU  & other &   & kpc        & mJy       & W/Hz     & W/Hz      &       \\
\hline
\endfirsthead
\caption{continued.}\\
\hline\hline
Name & Name  & z & Morphology & S$_c$(5.0)& Log P$_c$& Log P$_t$ & Notes \\
IAU  & other &   & kpc        & mJy       & W/Hz     & W/Hz      &       \\
\hline
\endhead
\hline
\endfoot
0034+25 & UGC367  & 0.0321 & FR I   &    10 &22.40 & 23.80 & N \\
0055+26 & N326    & 0.0472 & FR I   &    11 &22.75 & 25.40 &  N \\
0055+30 & N315    & 0.0167 & FR I   &   588 &23.57 & 24.24&   G \\
0104+32 & 3C31    & 0.0169 & FR I   &    92 &22.76 & 24.80 &  G \\
0106+13 & 3C33    & 0.0595 & FR II  &    24 &23.30 & 26.42 &  I \\
0116+31 & 4C31.04 & 0.0592 & C      &  1250 &23.42 & 25.46 & I, 1 \\
0120+33 & N507    & 0.0164 & FR I   &   1.4 &20.93 & 23.62 & no VLBI \\
0149+35 & N708    & 0.0160 & FR I   &   5   &21.46 & 23.32 & N \\ 
0206+35 & 4C35.03 & 0.0375 & FR I   &   106 &23.54 & 25.17 & G \\
0220+43 & 3C66B   & 0.0215 & FR I   &   182 &23.28 & 25.30 & G \\
0222+36 &         & 0.0327 &   C    &   140 &23.52 & 23.93 & G,5b \\
0258+35 & N1167   & 0.0160 & CSS    &$<$243&$<$23.14 & 24.35 & G,5b \\
0300+16 & 3C76.1  & 0.0328 & FR I   &  10   &21.38 & 24.09 & N \\
0326+39 &         & 0.0243 & FR I   &   78  &23.01 & 24.40 & I      \\
0331+39 & 4C39.12 & 0.0202 &   C    &   149 &23.11 & 24.20 & G \\
0356+10 & 3C98    & 0.0306 & FR II  &    9  &22.28 & 25.73 & N \\
0648+27 &         & 0.0409 &   C    &    58 &23.36 & 24.02 & G,5b \\
0708+32 &         & 0.0672 & FR I   &    15 &23.20 & 24.49 & N \\ 
0722+30 &         & 0.0191 & Spiral?&    51 &22.62 & 23.49 &  N \\
0755+37 & N2484   & 0.0413 & FR I   &   195 &23.82 & 25.36 & G \\
0800+24 &         & 0.0433 & FR I   &     3 &22.11 & 24.14 & no VLBI \\ 
0802+24 & 3C192   & 0.0597 & FR II  &     8 &22.83 & 26.01 & N \\
0828+32AB&4C32.15 & 0.0507 & FR II  &    3.3&22.29 & 25.41 & no VLBI \\
0836+29-I&4C29.30  & 0.0650 & FR I   &  63.0&  23.79 & 25.19 & N \\
0836+29-II&4C29.30  & 0.0790 & FR I    &  131& 24.28 & 25.42 & G \\
0838+32 & 4C32.26 & 0.0680 & FR I   &   7.5 &22.91 & 25.32 & N \\ 
0844+31 & IC2402  & 0.0675 & FR II  &   40  &23.56 & 25.53 & I   \\
0913+38 &         & 0.0711 & FR I   &$<$1.0&$<$22.08 & 24.99 & no VLBI \\
0915+32 &         & 0.0620 & FR I   &   8.0 &22.86 & 24.63 & N\\
0924+30 &         & 0.0266 & FR I   &$<$0.4&$<$20.80&24.44 & no VLBI \\
1003+35 & 3C236   & 0.0989 & FR II  &  400  &24.97 & 26.16 & I   \\ 
1037+30 & 4C30.19 & 0.0909 &   C    &$<$84&$<$24.22 & 25.36& I \\ 
1040+31 &         & 0.0360 &   C    &  55   &23.21 & 24.70 & I      \\
1101+38 & Mkn 421 & 0.0300 & BL Lac &  640  &24.11 & 24.39 & G \\
1102+30 &         & 0.0720 & FR I   &   26  &23.50& 25.07 & I       \\
1113+29 & 4C29.41 & 0.0489 & FR I   &   41  &23.27& 25.35 & N \\
1116+28 &         & 0.0667 & FR I   &   30  &23.50 & 25.05 & I, N* \\
1122+39 & N3665   & 0.0067 & FR I   &    6  &20.77 & 22.45 & N\\ 
1142+20 & 3C264   & 0.0206 & FR I   &  200  &23.28& 25.17 & G, 3 \\
1144+35 &         & 0.0630 & FR I   &  250  &24.37 & 24.49 & G, 7 \\
1204+24 &         & 0.0769 & FR I   &    8  &23.05 & 24.58 & N \\ 
1204+34 &         & 0.0788 & FR II? &   23  &23.53& 25.17 & I       \\ 
1217+29 & N4278   & 0.0021 &   C    &  162  &21.20& 21.75& G, 5 \\
1222+13 & 3C272.1 & 0.0037 & FR I   &   180 &21.74 & 23.59 & G \\
1228+12 & 3C274   & 0.0037 & FR I   &  4000 &23.30 & 25.38 & G \\
1243+26B&         & 0.0891 & FR I   &$<$1.8&$<$22.54 &25.23& no VLBI \\
1251+27 & 3C277.3 & 0.0857 & FR II  &    12 &23.33 & 26.05 & N \\
1254+27 & N4839   & 0.0246 & FR I   &   1.5 &21.31 & 23.45 & 9 \\
1256+28 & N4869   & 0.0224 & FR I   &    2.0&21.31 & 24.22 & no VLBI \\
1257+28 & N4874   & 0.0239 & FR I   &  1.1  &21.15 & 23.81 & 9 \\ 
1316+29 & 4C29.47 & 0.0728 & FR I   &   31  &23.58 & 25.59 & I       \\ 
1319+42 & 3C285   & 0.0797 & FR I/II&  6    &22.95& 25.92 & N \\
1321+31 & N5127   & 0.0161 & FR I   &  21   &22.09 & 24.32 & I       \\
1322+36 & N5141   & 0.0175 & FR I   & 150&   23.01 & 24.07 & G \\
1339+26 & UGC8669 & 0.0722 & FR I   &$<$55&$<$23.83 & 25.19& N \\
1346+26 & 4C26.42 & 0.0633 & FR I   &  53   &23.70 & 25.47 & I, N*, 8 \\
1347+28 &         & 0.0724 & FR I   &  4.8  &22.78 & 24.81 & no VLBI \\
1350+31 & 3C293   & 0.0452 & FR I   &$<$100&$<$23.67 &25.68& I, N*, 4 \\
1357+28 &         & 0.0629 & FR I   &  6.2  &21.97 & 24.81 & N \\
1414+11 & 3C296   & 0.0237 & FR I   &  77   &22.98 & 24.93 & I       \\
1422+26 &         & 0.0370 & FR I   &  25.0 &22.89 & 24.80 & I       \\
1430+25 &         & 0.0813 & FR I   &$<$1.0&$<$22.20&25.34& no VLBI \\ 
1441+26 &         & 0.0621 & FR I   &$<$0.7&$<$21.81 &24.78& no VLBI \\
1448+63 & 3C305   & 0.0410 & FR I   &   29  &23.05 & 25.44 & I \\
1502+26 & 3C310   & 0.0540 & FR I   &  80   &23.73 & 26.20 & I, N* \\
1512+30 &         & 0.0931 & C      &$<$0.4&$<$21.92 &24.75& no VLBI \\ 
1521+28 &         & 0.0825 & FR I   &   40  &23.81 & 25.40 & I       \\
1525+29 &         & 0.0653 & FR I   &  2.5  &22.40& 24.64 & no VLBI \\
1528+29 &         & 0.0843 & FR I   &  4.5  &22.89 & 25.10 & no VLBI \\
1529+24 & 3C321   & 0.0960 & FR II  &  30   &23.83 & 26.20 & I       \\
1549+20 & 3C326   & 0.0895 & FR II  &  3.5  &22.83 & 26.27 & no VLBI \\
1553+24 &         & 0.0426 & FR I   &  40   &23.22 & 23.92 & I      \\
1557+26 & IC4587  & 0.0442 &   C    &  31   &23.42 & 24.05 & I \\ 
1610+29 & N6086   & 0.0313 & FR I   &$<$6.0&$<$22.12 &23.85& no VLBI \\
1613+27 &         & 0.0647 & FR I   &  25   &23.40 & 24.65 & I      \\ 
1615+35B& N6109   & 0.0296 & FR I   &  28   &22.75 & 25.13 & N \\ 
1621+38 & N6137   & 0.0310 & FR I   &  50   &23.03 & 24.30 & I      \\ 
1626+39 & 3C338   & 0.0303 & FR I   &  105  &23.33 & 25.57 & G, 6 \\
1637+29 &         & 0.0875 & FR I   &  13   &23.38 & 25.17 & N \\ 
1652+39 & Mkn 501 & 0.0337 & BL-Lac & 1250  &24.51 & 24.36 & G, 2 \\
1657+32A&         & 0.0631 & FR I   &  2.5  &22.37 & 25.07 & no VLBI \\ 
1658+30 & 4C30.31 & 0.0351 & FR I   &  84   &23.36 & 24.63 & I       \\
1736+32 &         & 0.0741 & FR I   &   8   &23.01 & 24.89 & N \\ 
1752+32B&         & 0.0449 & FR I   &  12   &22.74 & 24.17 & N \\ 
1827+32A&         & 0.0659 & FR I   &  26   &23.43 & 24.88 & I      \\
1833+32 & 3C382   & 0.0586 & FR II  &  188  &24.18 & 26.04 & G \\
1842+45 & 3C388   & 0.0917 & FR II  &   62  &24.10 & 26.47 & I     \\
1845+79 & 3C390.3 & 0.0569 & FR II  &  330  &24.39 & 26.30 & G \\
1855+37 &         & 0.0552 &   C    &$<$100&$<$23.85 &24.75& I \\
2116+26 & N7052   & 0.0164 & FR I   &  47   &22.46 & 23.29 & I  \\
2212+13 & 3C442   & 0.0262 & FR I?  &  2.0  &21.49 & 25.21 & no VLBI \\
2229+39 & 3C449   & 0.0181 & FR I   &  37   &22.44 & 24.68 & I      \\
2236+35 & UGC12127& 0.0277 & FR I   &  8.0  &22.14 & 24.12 & N \\
2243+39 & 3C452   & 0.0811 & FR II  &  130  &24.30 & 26.64 & G \\
2335+26 & 3C465   & 0.0301 & FR I   &  270  &23.74 & 25.61 & G \\
\hline
\multicolumn{8}{l}{\scriptsize S$_c$(5.0) is the arcsecond core flux density at 5.0 GHz;}\\
\multicolumn{8}{l}{\scriptsize  
Log P$_c$ is the corresponding logarithm of the radio power;}\\
\multicolumn{8}{l}{\scriptsize 
Log P$_t$ is the logarithm of the total radio power at 408 MHz;}\\
\multicolumn{8}{l}{\scriptsize 
Morphology:
C = flat spectrum compact core, CSO = Compact Symmetric Object, 
CSS = Compact Steep Spectrum source,}\\
\multicolumn{8}{l}{\scriptsize 
FRI or II = Fanaroff type I or II;}\\
\multicolumn{8}{l}{\scriptsize  
Notes refer to the status of VLBI observations:}\\
\multicolumn{8}{l}{\scriptsize 
G: Giovannini et al. 2001;
I: Giovannini et al. 2005;
N, N* new data are given in this paper;} \\
\multicolumn{8}{l}{\scriptsize 
1: Giroletti et al. 2003a,b;
2: Giroletti et al. 2004, 2008
3: Lara et al. 2004
4: Beswick et al. 2004}\\
\multicolumn{8}{l}{\scriptsize 
5: Giroletti et al. 2005
5b: Giroletti et al. 2005b
6: Gentile et al. 2007;} \\
\multicolumn{8}{l}{\scriptsize 
7: Giovannini et al. 2007;
8: Liuzzo et al. 2009, submitted;
9: Liuzzo et al. 2009, in preparation;}\\
\multicolumn{8}{l}{\scriptsize 
For a reference to the extended structure see Fanti et al. 1987 for B2 sources
and Leahy et al. at the URL:
http://www.jb.man.ac.uk/atlas).} \\
\end{longtable}

\vfill\eject

\begin{table*}
\caption{Sources and Calibrators list\label{t.list}}
\centering
\begin{tabular}{lcccl} 
\hline\hline
Name &\multicolumn{2}{c}{Core Absolute Position}& phase & ATMCA \\ 
IAU  & RA$_{(J2000)}$ & DEC$_{J2000}$ & calibrator & calibrators \\ 
     & ($^{h\,m\,s}$) & ($^{\circ}\,'\,''$) \\ 
\hline
0034+25 & 00 37 05.488 & 25 41 56.331 & J0046+2456 & J0106+2539, J0057+3021, J0019+2602 \\ 
0055+26 & 00 58 22.630 & 26 51 58.689 & J0046+2456 & J0106+2539, J0057+3021, J0019+2602 \\ 
0149+35 & 01 52 46.458 & 36 09 06.494 & J0152+3716 & J0137+3309, J0205+3212  \\ 
0300+16 & 03 03 15.014 & 16 26 18.978 & J0305+1734 & J0256+1334, J0321+1221  \\ 
0356+10 & 03 58 54.437 & 10 26 02.778 & J0409+1217 & J0407+0742, J0345+1453 \\ 
0708+32B& 07 11 47.669 & 32 18 35.946 & J0714+3534 & J0646+3041 \\ 
0722+30 & ND           & ND           & J0736+2954 & J0646+3041 \\ 
0802+24 & 08 05 35.005 & 24 09 50.329 & J0802+2509 & J0813+2542,J0805+2106 \\ 
0836+29-I&08 40 02.352 & 29 49 02.529 & J0839+2850 & J0827+3525,J0852+2833,J0823+2928 \\ 
0838+32 & 08 41 13.097 & 32 24 59.712 & J0839+2850 & J0827+3525,J0852+2833,J0823+2928 \\ 
0915+32B& 09 18 59.406 & 31 51 40.636 & J0919+3324 & J0915+2933 \\ 
1113+29 & 11 16 34.619 & 29 15 17.121 & J1125+2610 & J1130+3031,J1103+3014,J1102+2757 \\ 
1116+28 & 11 18 59.366 & 27 54 07.003 & J1125+2610 & J1130+3031,J1103+3014,J1102+2757 \\ 
        & 11 18 59.366 & 27 54 07.002 & J1125+2610 & 1.6 GHz data \\ 
1122+39 & 11 24 43.624 & 38 45 46.278 & J1130+3815 & J1104+3812,J1108+4330 \\ 
1204+24 & 12 07 07.365 & 23 54 24.886 & J1209+2547 & J1150+2417,J1212+1925 \\ 
1251+27B& 12 54 12.007 & 27 37 33.961 & J1300+2830 & J1230+2518 \\ 
1319+42 & 13 21 17.867 & 42 35 14.989 & J1327+4326 & J1324+4048,J1254+4536 \\ 
1339+26B& ND           & ND           & J1342+2709 & J1327+2210 J1407+2827,J1329+3154 \\ 
1346+26 & 13 48 52.489 & 26 35 34.341 & J1342+2709 & J1327+2210,J1407+2827,J1329+3154 \\ 
        & 13 48 52.489 & 26 35 34.337 & J1350+3034 & 1.6 GHz data \\ 
1350+31 & 13 52 17.801 & 31 26 46.463 & J1350+3034 & 1.6 GHz data \\ 
1357+28 & 14 00 00.854 & 28 29 59.801 & J1342+2709 & J1327+2210, J1407+2827,J1329+3154 \\ 
1422+26 & 14 24 40.529 & 26 37 30.476 & J1436+2321 & 1.6 GHz data \\ 
1448+63 & 14 49 21.61  & 63 16 14.3   & J1441+6318 & 1.6 GHz data \\ 
1502+26 & 15 04 57.120 & 26 00 58.455 & J1453+2648 & J1522+3144,J1516+1932 \\ 
        & 15 04 57.109 & 26 00 58.473 & J1453+2648 & 1.6 GHz data \\ 
1529+24 & 15 31 43.47  & 24 04 19.1   & J1539+2744 & 1.6 GHz data \\ 
1557+26 & 15 59 51.614 & 25 56 26.319 & J1610+2414 & 1.6 GHz data \\ 
1613+27 & 16 15 31.360 & 27 26 57.335 & J1610+2414 & 1.6 GHz data \\ 
1615+35B& 16 17 40.537 & 35 00 15.201 & J1613+3412 & J1635+3808, J1642+2523 \\ 
1621+38 & 16 23 03.119 & 37 55 20.552 & J1640+3946 & 1.6 GHz data \\ 
1637+29 & 16 39 20.117 & 29 50 55.855 & J1653+3107 & J1635+3808, J1642+2523 \\ 
1736+32 & 17 38 35.770 & 32 56 01.524 & J1738+3224 & J1735+3616, J1753+2848 \\ 
1752+32B& 17 54 35.503 & 32 34 17.192 & J1748+3404 & J1735+3616, J1753+2848 \\ 
2236+35 & 22 38 29.421 & 35 19 46.872 & J2248+3718 & J2216+3518, J2236+2828 \\ 
\end{tabular}
\end{table*}

\clearpage
\vfill
\eject

\begin{table*}
\caption{VLBA Observations} \label{t.parameters}
\centering
\begin{tabular}{ccccrrc}
\hline\hline 
Epoch    & target & HPBW & noise & S$_{tot}$ & core/jet & notes \\
gg-mm-yy & & mas$\times$mas, $^\circ$ &mJy/beam &mJy &mJy & \\
\hline 
Epoch    & target & HPBW & noise & S$_{tot}$ & core/jet & notes \\
gg-mm-yy & & mas$\times$mas, $^\circ$ &mJy/beam &mJy &mJy & \\
2005-06-22&0034+25 & 3.3$\times$ 1.6 (-15) &0.1  & 3.6& 3.6& C \\
2005-06-22&0055+26 & 3.3$\times$ 1.6 (-14) &0.09 & 7.5& 7.5& C \\
2005-07-20&0149+35 & 3.0$\times$ 1.4 (0)   &0.08 & 3.2& 3.2& C \\ 
2005-07-20&0300+16 & 3.2$\times$ 1.7 (-1)  &0.1  & 6.1& 6.1& C \\
2005-07-20&0356+10 & 3.3$\times$ 1.6 (-4)  &0.12 & 2.2& 2.2& C \\
2005-07-20&0708+32B& 1.5$\times$ 1.5 (-)   &0.08 &10.5& 5.9& double:A \\ 
          &        &                       &     &    & 4.6& double:B \\
2005-07-20&0722+30 &                       & 0.1 & ND &ND  & - \\
2005-07-02&0802+24 & 3.5$\times$ 3.5 (-)   &0.05 & 3.6& 2.1& two-sided:C\\
          &        &                       &     &    & 1.5& two-sided:J\\
2005-07-02&0836+29-I&1.4$\times$ 1.4 (-)   &0.08 &47.8 &24.2& one-sided:C\\
          &         &                      &     &     & 5.8& one-sided:J1\\
          &         &                      &     &     &17.1& one-sided:J2\\
2005-07-02&0838+32 & 3.0$\times$ 1.7 (12)  &0.09 & 7.0 & 7.0& C \\
2005-06-17&0915+32 & 2.9$\times$ 1.6 (-6)  &0.1  &13.6 &13.6& C \\
2005-07-02&1113+29 & 4 $\times$  4 (-)     &0.06 &40.0 &36.8& one-sided:C\\
          &        &                       &     &     & 3.2& one-sided:J\\ 
2005-07-02&1116+28 & 3.3$\times$ 1.5 (5)   &0.09 &11.5 &11.5& C at 5 GHz\\
2003-08-07&1116+28 &$10.8\times5.5$ (28)   &0.13 & 9.5 &9.5 & C at 1.6 GHz\\
2005-06-17&1122+39 & 2.6$\times$ 1.8 (-1)  &0.10 &8.8  & 8.8& C \\
2005-06-17&1204+24 & 3.0$\times$ 1.7 (-4)  &0.09 &5.3  & 5.3& C \\
2005-06-17&1251+27B& 2$\times$ 2 (-)       &0.06 &10.4 & 6.2& two-sided:C\\
          &        &                       &     &     & 4.2& two-sided:J \\
2005-06-17&1319+42 & 2.3$\times$1.4  (0)   &0.1  &5.0  &3.0 & two-sided:C \\
          &        &                       &     &     &2.0 & two-sided:J \\
2005-06-27&1339+26 &                       &0.1  &ND   & ND  & - \\
2005-06-27&1346+26 & see Liuzzo et al.     &     &     &     & Z structure \\
2003-08-07&1350+31 &$12\times10$, (24)     &0.1  &235  &17.2 & at 1.6 GHz:C \\
          &        &                       &     &     & 2.5 & J          \\
          &        &                       &     &     & 2.0 & CJ         \\
          &        &                       &     &     &192&E1 see Sect.4.1.2\\
          &        &                       &     &     &11.9 & E2         \\
          &        &                       &     &     &11.2 & E3         \\
2005-06-27&1357+28 & 3$\times$ 2 (0)       &0.15 &5.6  & 3.8 & two-sided:C \\
          &        &                       &     &     & 1.8 & two-sided:J \\
2003-08-07&1422+26 &$10.1\times4.5$ (&0.15 &9.3  & 5.1 & one-sided:C? 1.6 GHz\\
          &        &                       &     &     & 4.2 & J?     \\
2003-08-30&1448+63 & 9.8$\times$ 4.5 (-5)  &0.15 &ND   &ND & 1.6 GHz,see text\\
2005-06-17&1502+26 & 2$\times$ 2 (-)     &0.06 & 7.8 & 0.5 &5GHz two-sided:C \\
          &        &                       &     &     & 3.1 & two-sided:E1 \\
          &        &                       &     &     & 4.2 & two-sided:W1 \\
2003-08-07&1502+26 &$10.2\times4.5$ (-20)  &0.2  &27.1 & 0.3 & C 1.6 GHz \\
          &        &                       &     &     & 6.0 & W \\
          &        &                       &     &     & 11.3 & E1 \\
          &        &                       &     &     & 4.1 & N \\
2003-08-07&1529+24 &11.0$\times$11.0 (-)   &0.1  &     &   & 1.6 GHz,see text\\
2003-08-30&1557+26 &$9.8\times4.5$ (2)     &0.12 &10.5 &10.5 & C 1.6 GHz \\
2003-08-30&1613+27 &$9.9\times4.5$ (3)     &0.12 & 9.3 & 9.3 & C 1.6 GHz \\
2005-06-27&1615+35B&2.8$\times$1.8 (-17)   &0.06 &23.6 &17.1 & one-sided:C \\
          &        &                       &     &     & 6.5 & one-sided:J \\
2003-08-07&1621+38 &7$\times$7 (-)     &0.09 &17.9&15.5 & 1.6GHz,one-sided:C \\
          &        &                       &     &      & 2.4 & one-sided:J\\
2005-06-27&1637+29 & 2.8$\times$ 1.8 (-12) &0.1  &7.8   & 7.8 & C \\
2005-06-22&1736+32 & 2.9$\times$ 1.2 (-32) &0.14 & 7.6  & 7.6 & C \\
2005-06-22&1752+32B& 2 $\times$ 2 (-)      &0.08 & 9.2  & 6.1 &two-sided:C \\
          &        &                       &     &      & 3.1 &two-sided:J \\
2005-06-22&2236+35 &3.3$\times$ 1.6 (-19)  &0.09 &8.8  & 8.8 & C \\
\hline \\
\multicolumn{7}{l}{\scriptsize 
In column 7 (notes) we indicate the component for which
flux density is reported in column 6 (core/jet). }\\
\multicolumn{7}{l}{\scriptsize 
C = core component;
j = jet. We also report the observing frequency when at 1.6 GHz, in all 
other cases the observations are at 5 GHz.} \\
\end{tabular}
\end{table*}

\clearpage

\begin{longtable}{ccccccccc}
\caption{Parameters for Observed Sources\label{tot.parameters}} \\
\hline\hline \\
Name & Type & Type & S$_{cVLBI}$ &S$_{totVLBI}$ &S$_{cVLA}$ & 
S$_{totVLBI}$/S$_{cVLA}$ & j/cj ratio & core \\
   & Kpc & pc & mJy & mJy & mJy & $\%$ & $\%$ &dominance \\
\hline
\endfirsthead
\caption{continued.}\\
\hline\hline \\
Name & Type & Type & S$_{cVLBI}$ &S$_{totVLBI}$ &S$_{cVLA}$ & 
S$_{totVLBI}$/S$_{cVLA}$ & j/cj ratio & core \\
   & Kpc & pc & mJy & mJy & mJy & $\%$ & $\%$ &dominance \\
\hline \\
\endhead
\hline \\
\endfoot
0034+25 &  FR I & C     &  3.6 &  3.6 &    10 & 36& ... & 1.10 \\
0055+26 &  FR I & C     &  7.5 &  7.5 &    11 & 68& ... & 0.26 \\
0055+30 &  FR I &two-sid&296   &477   &   588 & 81&   42& 8.71 \\
0104+32 &  FR I &one-sid& 72   & 90   &    92 & 98&$>$16& 0.60 \\
0106+13 &  FR II&two-sid& 11.0 & 30   &    24 &100&    2& 0.21 \\
0116+31 &  C    & CSO   & 20.0 &1250  &   --  &100& ... & 1.07 \\
0149+35 &  FR I & C     &  3.2 &  3.2 &   5   & 64& ... & 0.25 \\
0206+35 &  FR I &one-sid& 75   & 87   &   106 & 82&$>$14& 2.14 \\
0220+43 &  FR I &two-sid&105   &180   &   182 & 99&   10& 0.98 \\
0222+36 &    C  &two-sid&58.2  &120   &   140 & 86&...  &12.02 \\ 
0258+35 &  CSS  &one-sid& 7.4  &243   & $<$243&100& ... &$<$2.75 \\
0300+16 &  FR I & C     &  6.1 &  6.1 &  10   & 61& ... & 0.07 \\
0326+39 &  FR I &one-sid& 55.5 & 74   &   78  & 95&$>$25& 1.91 \\
0331+39 &    C  &one-sid& 90   &104   &   149 & 70&$>$12& 3.24 \\
0356+10 &  FR II& C     &  2.2 &  2.2 &    9  & 24& ... & 0.05 \\
0648+27 &    C  &one-sid&  4.4 & 12.8 &   58  & 22&$>$8 & 7.41 \\
0708+32B&  FR I &CSO?   &   ?  & 10.5 &    15 & 70& ... & 2.63 \\
0755+37 &  FR I &one-sid& 138  & 216  &   195 &100&$>$20& 3.16 \\
0802+24 &  FR II&two-sid&  2.1 &  3.6 &     8 & 45&  1  & 0.13 \\
0836+29-II&FR I &one-sid& 146  & 167  &    131&100&$>$20& 8.32 \\
0836+29-I& FR I &one-sid& 24.2 & 47.8 &   63&   76&$>$50& ...  \\
0838+32 &  FR I & C     &  7.0 &  7.0 &   7.5 & 93& ... & 0.41 \\
0844+31 &  FR II&two-sid& 27.0 & 32   &   40  & 80& 1   & 1.35 \\
0915+32 &  FR I & C     & 13.6 & 13.6 &   8.0 &100& ... & 0.98 \\
1003+35 &  FR II&one-sid&152.8 &466   &  400  &100&$>$45&14.13 \\
1037+30 &    C  & ND    &  ND  &$<$2  & $<$84 & --& ... &$<$10\\
1040+31 &    C  &two-sid& 37.6 & 48   &  55   & 87&    5& 2.00 \\
1101+38 & BL Lac&one-sid&335   & 448  &  640  & 70&$>$110&24.55\\
1102+30 &  FR I &two-sid& 14.3 & 17   &   26  & 65&1.7? & 2.29 \\
1113+29 &  FR I &one-sid& 36.8 & 40.0 &   41  & 98&$>$ 5& 0.89 \\
1116+28 &  FR I & C     & 11.5 & 11.5 &   30  & 38& ... & 2.34 \\
1122+39 &  FR I & C     &  8.8 &  8.8 &    6  &100& ... & 0.18 \\
1142+20 &  FR I &one-sid& 125  & 172  &  200  & 86&$>$24& 1.17 \\
1144+35 &  FR I &two-sid& 50.1 &  ?   &  --   & --& ... & .. \\
1204+24 &  FR I & C     &  5.3 &  5.3 &    8  & 66& ... & 4.57 \\
1204+34 &  FR II&one-sid&  33.7 & 39  &   23  &100&$>$2 & 2.09 \\
1217+29 &    C  &two-sid&  31.5 & 121 &  162  & 75&$>$2 &$<$2.75\\
1222+13 &  FR I &one-sid& 105   & 126 &   180 & 70&$>$10& 0.32 \\
1228+12 &  FR I &one-sid&450    &1600 &  4000 & 40&$>$150&0.91 \\
1251+27B&  FR II&two-sid&  6.2 & 10.4 &    12 & 87&   2 & 0.38 \\
1257+28 &  FR I &one-sid& 0.79 & 1.07 &  1.1  & 97&$>$2 & 0.06 \\
1316+29 &  FR I &two-sid&  5.3 & 23   &   31  & 74&   1 & 1.29 \\
1319+42 &  FR II&two-sid&  3.0 &  5.0 &  6    & 83&  2  & 0.19 \\
1321+31 &  FR I &two-sid&  3.7 & 11   &  21   & 52& 1?  & 0.26 \\
1322+36 &  FR I &one-sid&  45  & 60   &  150 &  40&$>$20& 3.09 \\
1339+26 &  FR I & ND    &  --  & --   & $<$55&  --& ... &$<$4.07\\
1346+26 &  FR I &two-sid&  4.1 & 44   &  53   & 83& ... & 2.04 \\
1350+31 &  FR I &two-sid& 14.4 & 23   & --    & --&  2  &$<$1.42\\
1357+28 &  FR I &two-sid&  3.8 &  5.6 &  6.2  & 90&  1? & 0.10 \\
1414+11 &  FR I &two-sid& 65.1 & 86   &  77   &100&  2  & 0.95 \\
1422+26 &  FR I &one-sid&  1.7 &  3   &  25.0 & 12&$>$20?& 0.81 \\
1448+63 &  FR I & C?    &  1.1 &  2   &   29  &  7& ... & 0.48 \\
1502+26 &  FR I &two-sid&  0.5 &  7.8 &  80   & 10&  1  & 0.78 \\
1521+28 &  FR I &one-sid& 39.5 & 44   &   40  &100&$>$17& 2.88 \\
1529+24 &  FR II& C     &  3.4 &  3   &  30   & 10& ... & 0.98 \\
1553+24 &  FR I &one-sid& 40.7 & 46   &  40   &100&$>$20& 6.17 \\
1557+26 &    C  &one-sid&  7.8 &  9   &  30   & 30&$>$ 4& 8.13 \\
1613+27 &  FR I & C     &  7.8 &  8   &  25   & 32& ... & 3.31 \\
1615+35B&  FR I &one-sid& 17.1 & 23.6 &  28   & 84&$>$15& 0.37 \\
1621+38 &  FR I &one-sid& 12.9 & 20   &  50   & 40&$>$ 5& 2.29 \\
1626+39 &  FR I &two-sid& 29   & 91   &  105  & 87&1.1  & 0.76 \\
1637+29 &  FR I & C     &  7.8 &  7.8 &  13   & 60& ... & 1.48 \\
1652+39 & BL-Lac&one-sid& 450  &1000  &  1250 & 80&$>$1250&64.54\\
1658+30 &  FR I &one-sid& 60.0 & 76   &  84   & 90&$>$8.5& 3.09 \\
1736+32 &  FR I & C     &  7.6 &  7.6 &   8   & 95& ... & 0.95 \\
1752+32B&  FR I &two-sid&  6.1 &  9.2 &  12   & 77&  2  & 1.41 \\
1827+32A&  FR I &one-sid&  5.9 & 15   &  26   & 58&$>$7.5& 2.51 \\
1833+32 &  FR II&one-sid& 105  & 131  &  188  & 70&$>$20& 2.75 \\
1842+45 &  FR II&one-sid& 34.0 & 52   &   62  & 84&$>$10& 1.23 \\
1845+79 &  FR II&one-sid& ??   & ??   &   330 & --&$>$100&3.02 \\ 
1855+37 &    C  & ND    &  ND  & $<$2 & $<$100& --& ... &$<$7.94\\
2116+26 &  FR I &two-sid&  ?   &  ?   &  47   & --& 1   & 2.63 \\
2229+39 &  FR I &  C    & 25.9 & 30   &  37   & 81& ... & 0.35 \\
2236+35 &  FR I & C     &  8.8 &  8.8 &  8.0  &100& ... & 0.39 \\
2243+39 &  FR II&two-sid& 18   & 125  &  130  & 96& 1.1 & 1.51 \\
2335+26 &  FR I &one-sid&168   & 234  &   270 & 87&$>$10& 1.82 \\
\hline \\
\multicolumn{9}{l}{\scriptsize 
VLBI data are at 5 GHz} \\
\end{longtable}
                                       

\begin{thebibliography}{}

\bibitem[Baum et al. 1988]{Baum1988} Baum, S.~A., Heckman, T.~M., Bridle, A.,
  van Breugel, W.~J.~M., \& Miley, G.~K.\ 1988, \apjs, 68, 643

\bibitem[Baum et al. 1990]{Baum1990} Baum, S.~A., Heckman, T., \& van Breugel,
  W.\ 1990, \apjs, 74, 389



\bibitem[Beswick et al. 2004]{Beswick2004} Beswick, R.~J., Peck, A.~B.,
  Taylor, G.~B., \& Giovannini, G.\ 2004, \mnras, 352, 49

\bibitem[Blandford et al. 1978]{bla78}
Blandford R. D., Icke V., 1978, MNRAS 185 527

\bibitem[Blanton et al. 2004]{Blanton2004} Blanton, E.~L., Sarazin, C.~L.,
  McNamara, B.~R., \& Clarke, T.~E.\ 2004, \apj, 612, 817


\bibitem[Burns et al. 1981]{Burns1981} Burns, J.~O., Gregory, S.~A., \&
  Holman, G.~D.\ 1981, \apj, 250, 450

\bibitem[Capetti et al. 2000]{Capetti2000} Capetti, A., de Ruiter, H.~R.,
  Fanti, R., Morganti, R., Parma, P., \& Ulrich, M.-H.\ 2000, \aap, 362, 871

\bibitem[Capetti et al.(1993)]{Capetti1993} Capetti, A., Morganti, R., Parma,
  P., \& Fanti, R.\ 1993, \aaps, 99, 407

\bibitem[Cotton 1993]{co93} Cotton, W.~D.\ 1993, \aj, 106, 1241 





\bibitem[de Ruiter et al. 1986]{deRuiter1986} de Ruiter, H.~R., Parma, P.,
  Fanti, C., \& Fanti, R.\ 1986, \aaps, 65, 111








\bibitem[Fanti et al. 1987]{Fanti1987} Fanti, C., Fanti, R., de Ruiter, H.~R.,
  \& Parma, P.\ 1987, \aaps, 69, 57

\bibitem[Fanti et al. 1986]{Fanti1986} Fanti, C., Fanti, R., de Ruiter, H.~R.,
  \& Parma, P.\ 1986, \aaps, 65, 145

\bibitem[Feretti et al. 1984]{fe84} 
Feretti L., Giovannini G., Gregorini L., Parma P., Zamorani G., 1984, A\&A 139, 55


\bibitem[Floyd et al. 2006]{Floyd2006} Floyd, D.~J.~E., Perlman, E., Leahy,
  J.~P., Beswick, R.~J., Jackson, N.~J., Sparks, W.~B., Axon, D.~J., \& O'Dea,
  C.~P.\ 2006, \apj, 639, 23

\bibitem[Gentile et al. 2007]{Gentile2007} Gentile, G., 
Rodr{\'{\i}}guez, C., Taylor, G.~B., Giovannini, G., Allen, S.~W., Lane, 
W.~M., \& Kassim, N.~E.\ 2007, \apj, 659, 225 

\bibitem[Giovannini et al. 1988]{g88} 
Giovannini G., Feretti L., Gregorini L., Parma P., 1988, A\&A 199, 73

\bibitem[Giovannini et al. 1990]{g90} 
Giovannini G., Feretti L., Comoretto G., 1990, \apj, 358, 159

\bibitem[Giovannini et al. 1994]{Giovannini1994} Giovannini, G., 
Feretti, L., Venturi, T., Lara, L., Marcaide, J., Rioja, M., Spangler, 
S.~R., \& Wehrle, A.~E.\ 1994, \apj, 435, 116 

\bibitem[Giovannini et al. 2001]{Giovannini2001} Giovannini, G., 
Cotton, W.~D., Feretti, L., Lara, L., \& Venturi, T.\ 2001, \apj, 552, 508 

\bibitem[Giovannini et al. 2005]{Giovannini2005} 
Giovannini G., Taylor G.B, Feretti L., Cotton W. D., Lara L., Venturi T., 2005, \apj 618, 635

\bibitem[Giovannini et al. 2007]{Giovannini2007} Giovannini, G., Giroletti, 
M., \& Taylor, G.~B.\ 2007, \aap, 474, 409 


\bibitem[Giroletti et al. 2003a]{gir03a}
Giroletti M., Giovannini G., Taylor G.B., Conway J.E., Lara L., Venturi T. 
2003a, A\&A 399, 889

\bibitem[Giroletti et al. 2003b]{gir03b} Giroletti, M.,Giovannini, G., Taylor, G.~B., 
\& Conway, J.~E.\ 2003b, New Astronomy Review, 47, 613 

\bibitem[Giroletti et al. 2004]{gir04} Giroletti, M., et  al.\ 2004, \apj, 600, 127 

\bibitem[Giroletti et al. 2008]{gir08} Giroletti, M., Giovannini, G., Cotton, W.~D., Taylor, G.~B., P{\'e}rez-Torres, M.~A., Chiaberge, M., \& Edwards, P.~G.\ 2008, \aap, 488, 905 

\bibitem[Giroletti et al. 2005a]{gir05a} Giroletti, M., Taylor, G.~B., \& Giovannini, G.\ 2005a, \apj, 622, 178 

\bibitem[Giroletti et al. 2005b]{gir05b} Giroletti, M., Giovannini, G., \& Taylor, G.~B.\ 2005b, \aap, 441, 89 

\bibitem[Giroletti et al. 2009]{gir09} Giroletti, M. et al. in preparation.

\bibitem[Gizani et al. 2002]{Gizani2002} Gizani, N.~A.~B., Garrett, M.~A., 
Morganti, R., Cohen, A., Kassim, N., Gonzalez-Serrano, J.~I., 
\& Leahy, J.~P.\ 2002, evn..conf, 163 

\bibitem[Gonzalez-Serrano et al. 1993]{go93} Gonzalez-Serrano, J.~I.,
  Carballo, R., \& Perez-Fournon, I.\ 1993, \aj, 105, 1710

\bibitem[Hardcastle et al. 1998]{Hardcastle1998} Hardcastle, M.~J., Alexander,
  P., Pooley, G.~G., \& Riley, J.~M.\ 1998, \mnras, 296, 445

\bibitem[Jamrozy et al. 2007]{jam07} Jamrozy, M., Konar, C., Saikia, D.~J., Stawarz, L., Mack, K.-H., 
\& Siemiginowska, A.\ 2007, \mnras, 378, 581 

\bibitem[Jetha et al. 2008]{Jetha2008} Jetha, N.~N., Hardcastle, M.~J.,
  Ponman, T.~J., \& Sakelliou, I.\ 2008, MNRAS in press (arXiv:0809.2534)


\bibitem[Lara et al. 2004]{Lara2004} Lara, L., Giovannini, G., Cotton, W.~D., Feretti, L., \& Venturi, T.\ 2004, \aap, 415, 905 

\bibitem[Leahy \& Williams 1984]{Leahy1984} Leahy, J.~P., \& Williams,
  A.~G.\ 1984, \mnras, 210, 929

\bibitem[Leahy \& Perley 1991]{Leahy1991} Leahy, J.~P., \& Perley,
R.~A.\ 1991, \aj, 102, 537

\bibitem[Ledlow et al. 2003]{Ledlow2003} Ledlow, M.~J., Voges, 
W., Owen, F.~N., \& Burns, J.~O.\ 2003, \aj, 126, 2740 

\bibitem[Liuzzo et al. 2009]{liu09a} Liuzzo et al. 2009, \aap, in press.
arXiv:0905.3453



\bibitem[Machalski 1998]{mac98}
Machalski, Astronomy and Astrophysics Supplement, 1998, v.128, p.153-178.

\bibitem[Massaro et al. 2009]{Massaro2009} Massaro, F., et al.\ 
2009, \apjl, 692, L123 







\bibitem[Murgia et al. 2001]{Murgia2001} Murgia, M., Parma, P., de Ruiter,
  H.~R., Bondi, M., Ekers, R.~D., Fanti, R., \& Fomalont, E.~B.\ 2001, \aap,
  380, 102


\bibitem[Parma et al. 1985]{Parma1985} Parma, P., Ekers, R.~D., \& Fanti,
  R.\ 1985, \aaps, 59, 511

\bibitem[Parma et al. 1986]{Parma1986} Parma, P., de Ruiter, H.~R., Fanti, C.,
  \& Fanti, R.\ 1986, \aaps, 64, 135



\bibitem[Pearson and Readhead 1988]{Pearson1988} 
Pearson T. J., Readhead A. C. S., 1988, \apj, 328, 114

\bibitem[Pearson et al. 1994]{pea94} Pearson, T.~J., Shepherd, M.~C., Taylor, G.~B., \& Myers, S.~T.\ 1994, Bulletin of the American Astronomical Society, 26, 1318 

\bibitem[Polatidis et al. 2003]{Polatidis1995} 
Polatidis A. G., Wilkinson P.N., Xu W., Readhead A.C.S., Pearson T. J., Taylor G. B., Vermeulen R. C., 1995, ApJS, 98, 1

\bibitem[Roche 
\& Eales 2000]{Roche2000} Roche, N., \& Eales, S.~A.\ 2000, \mnras, 317, 120 






\bibitem[Sandage 1972]{Sandage1972}
Sandage A., 1972, \apj, 178, 25



\bibitem[Taylor et al. 1994]{Taylor1994} 
Taylor G. B., Vermeulen R. C., Pearson T. J., Readhead A. C. S.,
 Henstock D. R., Browne I. W. A., Wilkinson P. N., 1994, ApJS 95, 345

\bibitem[Taylor 
\& Vermeulen 1997]{Taylor1997} Taylor, G.~B., \& Vermeulen, R.~C.\ 1997, \apjl, 485, L9 

\bibitem[Taylor et al. 2009]{Taylor2009}
Taylor G.B., Charlot P., Vermeulen R.C., Pradel N., 2009 \apj ~~in press; 
arXiv:0904.1879

\bibitem[Trussoni et al. 1997]{Trussoni1997} Trussoni, E., Massaglia, S.,
  Ferrari, R., Fanti, R., Feretti, L., Parma, P., \& Brinkmann, W.\ 1997, \aap,
  327, 27

\bibitem[Valentijn 1979]{val79} Valentijn, E.~A.\ 1979, \aap, 78, 367 


\bibitem[Vall\`ee 1982]{Vallee1982} Vall\`ee, J.~P.\ 1982, \aj, 87, 486

\bibitem[van Breugel \& Fomalont 1984]{breu84} van Breugel, W., \&
  Fomalont, E.~B.\ 1984, \apjl, 282, L55

\bibitem[van Breugel et al. 1986]{breu86} van Breugel, W.~J.~M., Heckman, T.~M., Miley, G.~K., 
\& Filippenko, A.~V.\ 1986, \apj, 311, 58 


\bibitem[van Breugel et al. 1985]{breu85} van Breugel W., Miley G.,
  Heckman T., Butcher H., Bridle A., 1985, \apj, 290, 496


\bibitem[Venturi et al. 1995]{Venturi1995} Venturi, T., Castaldini, C.,
  Cotton, W.~D., Feretti, L., Giovannini, G., Lara, L., Marcaide, J.~M., \&
  Wehrle, A.~E.\ 1995, \apj, 454, 735








\end{thebibliography}
\end{document}